\documentclass[aps,superscriptaddress,showkeys]{revtex4}
\usepackage[dvips]{graphicx}

\begin{document}

\title{Compact Stars for Undergraduates}

\author{Irina Sagert}
\email{sagert@astro.uni-frankfurt.de}

\author{Matthias Hempel}
\email{hempel@astro.uni-frankfurt.de}

\author{Carsten Greiner}
\email{carsten.greiner@th.physik.uni-frankfurt.de}

\author{J\"urgen Schaffner--Bielich}
\email{schaffner@astro.uni-frankfurt.de}

\affiliation{
Institut f\"ur Theoretische Physik,
J. W. Goethe Universit\"at,
Max--von--Laue--Stra\ss{}e 1,
D--60438 Frankfurt am Main,
Germany}

\begin{abstract}
 We report on an undergraduate student project initiated in the summer
 semester of 2004 with the aim to establish equations of state for
 white dwarfs and neutron stars for computing mass--radius relations as
 well as corresponding maximum masses. First, white dwarfs are
 described by a Fermi gas model of degenerate electrons and neutrons
 and effects from general relativity are examined. For neutron star
 matter, the influence of a finite fraction of protons and electrons
 and of strong nucleon--nucleon interactions are studied. The
 nucleon--nucleon interactions are introduced within a Hartree--Fock
 scheme using a Skyrme--type interaction. Finally, masses and radii of
 neutron stars are computed for given central pressure.
\end{abstract}


\maketitle

\section{Introduction}

Compact stars, i.e.\ white dwarfs and neutron stars, are the final
stages in the evolution of ordinary stars. After hydrogen and helium
burning, the helium reservoir in the star's core is burned up to carbon
and oxygen. The nuclear processes in the core will stop and the star's
temperature will decrease. Consequently, the star shrinks and the
pressure in the core increases.

As long as the star's mass is above a certain value, it will be able to
initialise new fusion processes to heavier elements. The smaller the
mass, the more the core has to be contracted to produce the required
heat for the next burning process to start. If the star's initial mass
is below eight solar masses, the gravitational pressure is too weak to
reach the required density and temperature to initialise carbon fusion.
Due to the high core temperature, the outer regions of the star swell
and are blown away by stellar winds. What remains is a compact core
mainly composed of carbon and oxygen. The interior consists of a
degenerate electron gas which, as we will discuss, is responsible for
the intrinsic high pressure. The compact remnant starts to emit his
thermal energy and a white dwarf is born.

As there are no fusion processes taking place in the interior of the
white dwarf, a different force than thermal pressure is needed to keep
the star in hydrostatic equilibrium. Due to the Pauli principle, two
fermions cannot occupy the same quantum state, hence, more than two
electrons (with different spin) cannot occupy the same phase space. With
increasing density, the electrons fill up the phase space from the
lowest energy state, i.e.\ the smallest momentum. Consequently, the
remaining electrons sit in physical states with increasing momentum. The
resulting large velocity leads to an adequate pressure of the electrons
--- the degeneracy pressure --- counterbalancing gravity's pull
and stabilising the white dwarf.

The electrons are the first particles to become degenerate in dense
matter due to their small mass. White dwarfs consist also of carbon
and oxygen, but the contribution of nuclei to the pressure is negligible
for the density region of interest here. The gravitational force
increases with mass. Consequently, a larger degeneracy pressure, i.e.\ a
greater density, is needed for stability for more massive white dwarfs.
There is a mass limit for white dwarfs beyond which even the degenerate
electron gas cannot prevent the star from collapsing. This mass limit is
around 1.4 solar masses --- the famous Chandrasekhar mass. For greater
masses, the white dwarf collapses to a neutron star or a black hole.
The radii of white dwarfs are in the range of 10,000 km --- about the
size of our planet earth.

Similarly, a neutron star is stabilised by the degeneracy pressure also.
The difference is that for neutron stars the degeneracy pressure
originates mainly from neutrons, not from electrons.

If the initial mass of an ordinary star exceeds eight solar masses,
carbon and oxygen burning starts in the core. Around the core is a layer
of helium and a layer of hydrogen, both taking part in fusion processes.
As long as they produce the required temperature to keep the star in
hydrostatic equilibrium, new burning processes to heavier elements will
take place in the core while the fusion of lighter elements will
continue in the outer shells. Once iron is produced in the core, there
is no burning process left to generate energy as the fusion to heavier
elements requires energy to be put in. So the mass of the iron core
increase while its radius decrease as there is less thermal pressure to
work against gravity. At some critical mass, the iron core collapses.
The density increases so much that the electrons are captured by nuclei
and combine with protons to form neutrons. Therefore, the atomic nuclei
become more and more neutron--rich with increasing density. At some
critical density, the nuclei are not able to bind all the neutrons any
more, neutrons start to drop out and form a neutron liquid which
surrounds the atomic nuclei. For even larger density there will be just
a dense and incompressible core of neutrons with a small fraction of
electrons and protons. Neutrons in neutron stars are then degenerate
like electrons in white dwarfs.

There will be an outgoing shock wave generated by the proto--neutron star
due to the high incompressibility of neutron star matter. The falling
outer layers of the ordinary star bounce and move outwards interacting
with the other still collapsing layers and generate an overall outward
expansion --- a core--collapse supernova arises. The neutron rich
remnant of the core collapse will become a neutron star as long as his
mass is less than about two to three solar masses, otherwise it proceeds
to shrink and becomes finally a black hole. The radii of neutron stars
are typically 10 km --- the size of the city of Frankfurt.

During the summer semester of 2004, the two last coauthors organised a
student project about compact stars on which we will give an extended
report in the following. We derive the equations of state (EoS) for
white dwarfs and neutron stars, respectively, where we will cover the
Newtonian and the relativistic cases for both kinds of compact stars.
These EoSs are used to determine numerically the masses and radii for a
given central pressure. In the first chapter we give a short
description of the properties of white dwarfs and neutron stars. We then
proceed in the second chapter with calculations of the EoS for white
dwarfs. We assume a Fermi gas model of degenerate electrons and outline
two different ways which we employed for deriving the equations of
state, repeat the basic structure equations for stars as well as the
corrections from general relativity (the so called
Tolman--Oppenheimer--Volkoff (TOV) equation) and discuss the Chandrasekhar
mass, which is the famous mass limit for white dwarfs. From the third
chapter on we will concentrate on neutron stars. Again we will have two
different approaches for calculating the equation of state for neutron
stars. After describing a pure neutron star, we will discuss the effect
of the presence of protons and electrons which will appear in
$\beta$--equilibrium. Furthermore, we will introduce a simple effective
model for incorporating the nucleon--nucleon interactions. We will
compare our results to the recent work by Reddy and Silbar
\cite{Silbar04} who are using different numerical methods and another
model for the nucleon--nucleon interactions. We investigate whether the
two model descriptions satisfy causality. Finally, we summarise our
findings and give a short outline of modern developments in neutron star
physics.

\section{White Dwarfs}
\label{whitedwarfs}

\subsection{Definitions}

\subsubsection{Structure equations}
\label{Structure equations}

There are two forces acting on the star, one of
them is gravitation and the second one arises from the pressure
(thermal pressure for normal stars and degeneracy pressure for white
dwarfs and neutron stars). The pressure and force are related in the
standard way:
\begin{eqnarray} 
{dp} & = & \frac{dF}{A} = \frac{dF}{4\pi r^2}\quad ,
\end{eqnarray}
where $p$ denotes the pressure, $F$ the force, and $A$ is the area where
the force takes effect. For the gravitational force one has for a
spherical symmetric system: 
\begin{eqnarray}
{dF} & = & - \frac{G {dm}\cdot m(r)} {r^2}\quad ,
\end{eqnarray}
with $r$ being the radial distance of a spherical star and 
\begin{eqnarray}
{dm} & = & \rho(r) {dV} = \rho(r) 4 \pi r^2 {dr}\quad .
\label{eq:dm}
\end{eqnarray}
Equating both expressions gives the first basic structure equation for
stars in general: 
\begin{eqnarray}
\frac{dp}{dr} & = & -\frac{G \rho(r) {m(r)}} {r^2}\quad .
\label{eq:pressure}
\end{eqnarray} 
Here $G$ is Newton's Gravitational constant, $\rho$ is the mass
density, $m$ is the mass up to radius r and $V$ is the volume. Again
using eq.~(\ref{eq:dm}), one has also the equation: 
\begin{eqnarray}
\frac{dm}{dr} & = & \rho(r) 4 \pi r^2 \quad.
\label{eq:ddm}
\end{eqnarray}
We can rewrite the mass density in terms of the energy density 
$\epsilon$: 
\begin{eqnarray}
\rho(r) & = & \frac{\epsilon(r)} {c^2}\quad ,
\end{eqnarray}
where $c$ is the speed of light. With these definitions one obtains
structure equations which describe how the mass and the pressure of a
star change with radius: 
\begin{eqnarray}
\frac{dm}{dr} & = & \frac{4 \pi r^2 \epsilon(r)} {c^2}
\label{eq:structure_M} \\
\frac{dp}{dr} & = & - \frac{G \epsilon(r) m(r)} {c^2 r^2}\quad .
\label{eq:structure_p}
\end{eqnarray}
Here one has to solve two coupled differential equations.
Note that while ${dm}/{dr}$ is positive, ${dp}/{dr}$ must always be
negative. Starting with certain positive values for $m$ and $p$ in a
small central region of the star the mass will increase while the
pressure decreases eventually reaching zero. One sets $m(r=0)=0$ and,
in addition, one has to specify some "initial" central pressure
${p(r=0)=p_0}$ in order to solve for eqs.~(\ref{eq:structure_M}) and
(\ref{eq:structure_p}). The behaviour of $m$ and $p$ as a function of
the radius will become important for our numerical computations.

\subsubsection{Equations of state}

From the last section, we arrived at two coupled differential equations
for the mass and the pressure in a star where both depend on the energy
density $\epsilon$. One needs to fix now the relation ${p(\epsilon)}$
between the pressure and the energy density in addition. Matter in white
dwarfs can be treated as an ideal Fermi gas of degenerate electrons. For
electrically neutral matter an adequate amount of protons to the
electron gas has to be added. Furthermore, the protons combine with
neutrons and form nuclei which will be also taken into account.

The distribution of fermions as an ideal gas in equilibrium in
dependence of their energy is described by the Fermi--Dirac--Statistic:
\begin{eqnarray}
{f(E)} & = & \frac{1} {exp[(E - \mu)/k_BT]+1} \quad ,
\end{eqnarray}
where $E$ is the energy, $\mu$ is the chemical potential, $k_B$ is the
Boltzmann constant and $T$ is the temperature. In kinetic theory we find
the following correlation between the distribution function ${f}$ and
the number density $n_i=dN_i/dV$ in phase space for a certain particle
species $i$: 
\begin{eqnarray}
\frac{dn_i}{d^3k} & = & \frac{g}{(2 \pi \hbar)^3}f \quad ,
\label{eq:number_phase_space}
\end{eqnarray}
$(2\pi \hbar)^3$ is the "unit" volume of a cell in phase space and $g$
is the number of states of a particle with a given value of momentum
$k$. For electrons $g$ equals 2. The number density of species $i$ is
given by: 
\begin{eqnarray}
{n_i} & = & \int dn_i = \int \frac{g}{(2 \pi \hbar)^3} f d^3k \quad .
\end{eqnarray}
For degenerate fermions the temperature can be set to zero (as $(\mu-m_e c^2)
\gg k_BT$), accordingly $\mu / k_BT$ goes to infinity. To be added to
the system, a particle must have an energy equal to the Fermi energy of
the system as all the other energy levels are already filled. This
description matches exactly the definition of the chemical potential,
and so we can set here $\mu$= ${E_F}$. With all these assumptions we can
write the distribution function as:
\begin{eqnarray}
f(E) & = & \left\{ \begin{array}{ccccc} 1 \quad \mbox{ for } \quad E \leq E_F\\ 
0 \quad \mbox{ for } \quad E > E_F \end{array} \right\}\quad . 
\end{eqnarray}
If we ignore all electrostatic interactions, we can write for the number
density of the degenerate electron gas: 
\begin{eqnarray}
{n_e} & = & \int_0^{k_F} \frac{2}{(2 \pi \hbar)^3} {d^3k} = \frac{8
 \pi}{(2 \pi \hbar)^3} \int_0^{k_F} k^2 dk = \frac{k_F^3}{3 \pi^2
 \hbar^3}\quad . 
\label{eq:nelectrons}
\end{eqnarray}
As the star should be electrically neutral, each electron is neutralised
by a proton. If we assume that the white dwarf is predominantly made of
$^{12}{\rm C}$ or $^{16}{\rm O}$, $A/Z = 2$. The proton and neutrons
masses are both much larger than the electron's one, so when we calculate
the mass density in the white dwarf we can neglect the mass of the
electrons and just concentrate on the nucleon mass $m_N$. The mass
density in terms of $m_N$ is: 
\begin{eqnarray}
\rho = n \cdot {m_N} \cdot \frac{A}{Z}\quad ,
\label{eq:rho}
\end{eqnarray}
where $n = n_e$. With this equation and the dependence of number density
on $k_F$ we find: 
\begin{equation}
 {k_F} = \hbar\left(\frac{3\pi^2\rho}{m_N}\frac{Z}{A}\right)^{1/3}\quad .
\label{eq:kF}
\end{equation}
As the momentum of the nucleons is negligible compared to their rest
mass, the nucleons do not give a significant contribution to the
pressure at zero temperature. On the other hand, the electrons behave as
a degenerate gas, and have large velocities which give the dominant
contribution to the pressure. The energy density can be divided in two
components, one term coming from nucleons and the other one coming from
electrons, whereas the contribution from nucleons is dominating. The
complete energy density can be written as \cite{Shapiro_book}:
\begin{eqnarray}
\epsilon & = & n m_N \frac{A}{Z} c^2 + \epsilon_{\rm elec}(k_F) \quad.
\label{eq:epsilon}
\end{eqnarray}
where $\epsilon_{\rm elec}(k_F)$ is the energy density of electrons.
With the electron energy
\begin{eqnarray}
E(k) & = & \sqrt{{k^2} c^2 + {m_e^2} c^4}
\label{eq:EF}
\end{eqnarray}
we can write $\epsilon_{elec}$ in the following way:
\begin{eqnarray}
 \epsilon_{\rm elec}(k_F) & = & \frac{8 \pi}{(2 \pi \hbar)^3}
 \int_0^{k_F} E(k) k^2 d k 
 \nonumber \\
 & = &\frac{8 \pi}{(2 \pi \hbar)^3} \int_0^{k_F} (k^2 c^2
 + m_e^2 c^4)^{1/2} k^2 d k 
 \nonumber \\
 & = & \epsilon_0 \int_0^{k_F/m_e c} (u^2 + 1)^{1/2} u^2 d u 
 \nonumber \\
 & = & \frac{\epsilon_0}{8} \left[ (2 x^3 + x)(1 + x^2)^{1/2}
 - \sinh^{-1}(x) \right] 
 \label{eq:electroneps} \ 
\end{eqnarray}
with 
\begin{eqnarray}
 \epsilon_0 = \frac{m_e^4 c^5}{\pi^2 \hbar^3}
\label{eq:epsilon0}
\end{eqnarray}
and 
\begin{equation}
x = k_F / m_e c .
\label{eq:x}
\end{equation}
The factor $\epsilon_0$ carries the dimension of an energy density
(${\rm dyne/{cm^2}}$). The pressure of a system with an isotropic
distribution of momenta is given by: 
\begin{eqnarray}
p & = & \frac{1}{3} \frac{8 \pi}{(2 \pi \hbar)^3} \int_0^{k_F} k v k^2 d k
\label{eq:p_fermi} 
\end{eqnarray}
where the velocity $v = {kc^2}/{E}$ and the factor ${1}/{3}$ comes from
isotropy. For the electrons we get: 
\begin{eqnarray}
p(k_F) & = & \frac{1}{3} \frac{8 \pi}{(2 \pi \hbar)^3} \int_0^{k_F}
\frac{k^2c^2}{E(k)} k^2 dk
\nonumber\\
& = & \frac{8 \pi}{3 (2 \pi \hbar)^3} 
 \int_0^{k_F} (k^2 c^2 + m_e^2 c^4)^{-1/2} c^2 k^4 d k 
 \nonumber \\
 & = & \frac{\epsilon_0}{3} \int_0^{k_F/m_e c} 
 (u^2 + 1)^{-1/2} u^4 d u 
 \nonumber \\
 & = & \frac{\epsilon_0}{24} \left[ (2 x^3 -3 x)(1 + x^2)^{1/2}
 +3 \sinh^{-1}(x) \right].
 \label{eq:electronpres} \
\end{eqnarray}
The energy density is dominated by the mass density of nucleons
while the electrons contribute to most of the pressure. 

We want to arrive at an equation of the form $p = p(\epsilon)$. In the
following, let us consider the two extreme cases: $x \ll 1$ and $x \gg
1$, i.e.\ $k_F \ll m_ec$ and $k_F \gg m_ec$, respectively. In the former
case, the kinetic energy is much smaller than the rest mass of the
electrons, i.e.\ that is the non--relativistic case. Considering the
equation for the pressure for $k_F \ll m_ec$ one finds that
\begin{eqnarray}
p(k_F)=\frac{\epsilon_0}{3} \int_0^{k_F/m_e c} (u^2 + 1)^{-1/2} u^4 d u \approx
\frac{\epsilon_0}{3} \int_0^{k_F/m_e c} u^4 du = \frac{\epsilon_0}{15}
\left(\frac{k_F}{m_e c}\right)^5 = \frac{\hbar^2}{15 \pi^2 m_e}
\left(\frac{3 \pi^2 \rho Z}{m_NA }\right)^{5/3}\quad . 
\end{eqnarray}
With the assumption $\epsilon=\rho\cdot c^2$ one arrives at the
following equation of state (EoS) in the non--relativistic limit:
\begin{eqnarray}
p& \approx & K_{\rm non-rel} \epsilon^{5/3} \label{eq:polytropnonrel}
\end{eqnarray}
with
\begin{eqnarray}
K_{\rm non-rel} & = & \frac{\hbar^2}{15 \pi^2 m_e} \left(\frac{3 \pi^2
 Z}{m_N c^2A }\right)^{5/3}\quad . 
\end{eqnarray}
For the relativistic case ($k_F \gg m_e$) we arrive at:
\begin{eqnarray}
 p(\epsilon)& \approx & K_{\rm rel} \, \epsilon^{4/3}\quad ,
 \label{eq:polytroprel}
\end{eqnarray}
where
\begin{eqnarray}
 K_{\rm rel} = \frac{\hbar c}{12 \pi^2}
 \left( \frac{3 \pi^2 Z}{m_N c^2A } \right)^{4/3}\quad .
\label{eq:K_rel}
\end{eqnarray}
The relation of the form 
\begin{equation}
p = K\epsilon^\gamma
\label{eq:polytrop}
\end{equation}
is called a "polytropic" equation of state. With the two EoSs for the
relativistic and for the non--relativistic case we are in the position to
calculate the mass and the radius of a white dwarf with a given central
energy density by inserting these relations into the structure equations
(\ref{eq:structure_M}) and (\ref{eq:structure_p}). A discussion on
general forms of polytropic EoSs can be found in \cite{Herrera04}.

Besides the non--relativistic and the relativistic polytropic EoS there
exists also a third case, which becomes important for later
considerations --- the ultra--relativistic EoS. Here one assumes
for simplicity a relativistic pure electron Fermi gas with the total
energy density:
\begin{equation}
\epsilon = \epsilon_{elec}(k_F).
\end{equation} 
Just like in eq.~(\ref{eq:electroneps}) one can calculate the energy
density using 
\begin{equation}
 \epsilon(k_F) = \epsilon_0 \int_0^{k_F/m_e c} (u^2 + 1)^{1/2} u^2 d u, 
\end{equation}
getting for the relativistic case
\begin{equation}
\epsilon = \frac{\epsilon_0}{4} x^4 \quad .
\end{equation}
From eq.~(\ref{eq:electronpres})
\begin{equation}
p(k_F) = \frac{\epsilon_0}{3} \int_0^{k_F/m_e c}(u^2 + 1)^{-1/2} u^4 d u 
\end{equation}
one finds for $k_F\gg m_e$:
\begin{equation}
p = \frac{\epsilon_0}{12} x^4.
\end{equation}
Therefore, one arrives at the general solution for an ultra--relativistic
gas: 
\begin{equation}
p = \frac{1}{3} \epsilon \quad .
\label{eq:ultra}
\end{equation}
This relation is also very important for neutron stars as a limiting
case of their relativistic EoS.

\subsubsection{Chandrasekhar mass}
\label{chandra}

We will derive in this section the maximum mass for white dwarfs for a
polytropic EoS as found by Chandrasekhar in 1931 \cite{Chandra31},
following \cite{Weinberg72,Shapiro_book}. Note, that already in 1932 Landau
found with simple arguments a mass limit for white dwarfs and neutron
stars \cite{Landau32} (see also the discussion in e.g.\ 
\cite{Shapiro_book}).

By straight--forward algebra the structure equations
(\ref{eq:structure_M}) and (\ref{eq:structure_p}) can be combined to
give:
\begin{eqnarray}
\frac{1}{r^2}\frac{d}{dr} \left(\frac{r^2}{\rho}\frac{dp}{dr}\right) & =
& - 4 \pi G \rho \quad . 
\label{eq:laneemden1}
\end{eqnarray}
The parameter $\gamma$ in the polytropic equation is usually rewritten as 
$\gamma = 1+ \frac{1}{n}$ where $n$ is the polytropic index.

Given the equation of state $p = p(\rho)$ in the polytropic form
(see eq.~(\ref{eq:polytrop}))
\begin{eqnarray}
p = K\epsilon^ \gamma = K \rho^ \gamma c^{2\gamma}
\end{eqnarray}
one can obtain $\rho(r)$ by solving eq.~(\ref{eq:laneemden1}) with the
initial conditions: 
\begin{equation}
\rho(r=0) = \rho_0 \neq 0
\end{equation}
and
\begin{eqnarray}
\left.\frac{d\rho}{dr}\right|_{(r=0)}=0
\label{eq:prime_rho}
\end{eqnarray}
which follows from the condition ${m(r=0)=0}$ and
eq.~(\ref{eq:structure_p}). Equation~(\ref{eq:laneemden1}) can be
transformed into a dimensionless form with the following substitutions:
\begin{equation}
r =  a \xi \quad \mbox{\rm with } \quad 
a \equiv \left(\frac{(n+1)K\rho_0^{(1-n)/n}}{4 \pi G}\right)^{1/2}
c^{(n+1)/n}  
\label{eq:r}
\end{equation}
and
\begin{equation}
\rho(r) =  \rho_0 \theta^n \quad \mbox{\rm with } \quad 
\theta = \theta(r) \quad .
\label{eq:rho0}
\end{equation}
Here $\theta$ and $\xi$ correspond to a dimensionless density and
radius, respectively, and $a$ is a constant scale factor. 
With these definitions eq.~(\ref{eq:laneemden1}) translates to:
\begin{eqnarray}
\frac{1}{\xi^2}\frac{d}{d\xi}\left(\xi^2 \frac{d\theta}{d\xi}\right) & =
& - \theta^n. 
\label{eq:laneemden2}
\end{eqnarray}
Equation (\ref{eq:laneemden2}) is called the Lane--Emden equation for a
given polytropic index $n$. From eq.~(\ref{eq:rho0}) one realizes that
\begin{eqnarray}
\theta(r=0)=1.
\label{eq:theta}
\end{eqnarray}
Furthermore, with eq.~(\ref{eq:prime_rho}) the density distribution at the
center of the star obeys:
\begin{eqnarray}
\left.\frac{d\theta}{dr}\right|_{(r=0)}=0.
\label{eq:frac_theta}
\end{eqnarray}
With the boundary conditions, eqs.~(\ref{eq:frac_theta}) and
(\ref{eq:theta}), one can integrate eq.~(\ref{eq:laneemden2})
numerically. One finds for $n < 5$ and $\gamma > 6/5$, respectively,
that the solutions decrease monotonically with radius and have a zero
for a $\xi = \xi_1$, i.e.\ $\theta(\xi_1) = 0$ or $\rho(r_1 =a \xi_1) =
0$, respectively. Hence, the radius of the star is given by R $= r_1 =
a\xi_1$ which translates to
\begin{eqnarray}
R & = & \left(\frac{(n+1)K\rho_0^{(1-n)/n}}{4 \pi G}\right)^{1/2}
c^{(n+1)/n} \xi_1. \label{eq:chr} 
\end{eqnarray}
Using eqs.~(\ref{eq:rho0}) and (\ref{eq:r}) the total mass $M$ of the star
can be calculated as follows: 
\begin{eqnarray} 
M & = & \int_0^R 4\pi r^2 \rho dr
\nonumber\\
& = & 4 \pi a^3 \rho_0 \int_0^{\xi_1} \xi^2 \theta^n d\xi
\nonumber\\
& = & 4 \pi c^{(3n+3)/n} \left( \frac{(n+1)K}{4 \pi G}\right)^{3/2}
\rho_0^{(3-n)/2n}\xi_1^2 | \theta^{\prime} (\xi_1)|.
\end{eqnarray} 
For $\rho_0$ we can use eq.~(\ref{eq:chr}) and get:
\begin{eqnarray}
M & = & 4 \pi c^{(2n+2)/(n-1)} \left( \frac{(n+1)K}{4 \pi
 G}\right)^{n/(n-1)} \xi_1^{(n-3)/(1-n)}\xi_1^2 | \theta^{\prime} 
(\xi_1)| R^{(3-n)/(1-n)}\quad .
\label{eq:masslong}
\end{eqnarray}
The interesting case is the high density limit, i.e.\ the relativistic
case with $\gamma = \frac{4}{3}$. For this case it follows
\cite{Weinberg72}: 
\begin{eqnarray}
\gamma = \frac{4}{3}: \quad {\rm n\ = 3}, \quad {\rm \xi_1 = 6.89685},
\quad {\rm \xi_1^2 | \theta^{\prime} 
(\xi_1)| = 2.01824}\quad .
\end{eqnarray}
For $K$ we take eq.~(\ref{eq:K_rel}) and set $\eta = A/Z$. Using these values
in eq.~(\ref{eq:masslong}), we get the following expressions for the
mass and the radius in the relativistic case (see \cite{Weinberg72}).
For the radius:
\begin{eqnarray}
R = \frac{1}{2} (3\pi)^{1/2}
(6.89685)\left(\frac{\hbar^{3/2}}{c^{1/2}G^{1/2}m_e m_N \eta}\right) 
\left(\frac{\rho_{\rm crit}}{\rho_0}\right)^{1/3}
\label{eq:chandra_radius}
\end{eqnarray}
with 
\begin{eqnarray}
\rho_{\rm crit} = \frac{m_N \eta {m_e}^3 c^3}{3 \pi^2 \hbar^3}
\end{eqnarray}
and for the mass:
\begin{eqnarray}
M = \frac{1}{2} (3\pi)^{1/2} (2.01824) \left(\frac{\hbar
 c}{G}\right)^{3/2} \left(\frac{1}{m_N \eta}\right)^2 
\end{eqnarray}
which becomes after plugging in all the values \cite{PDBook},
\begin{eqnarray}
M = 1.4312 \left(\frac{2}{\eta}\right)^2 {\rm M_\odot}\quad . 
\label{eq:chandrasekhar}
\end{eqnarray} 
Here and for the later calculations we chose for $m_N$ the neutron mass.
For a white dwarf which consists mainly of $^{12}{\rm C}$ it is better
to take the atomic mass unit (again from \cite{PDBook}), which gives a
maximum mass of 1.4559 solar masses. It is remarkable that in the
relativistic case the mass does not depend on the radius $R$ or the
central density $\rho_0$ anymore. As the density in the white dwarf
increases, the electrons become more relativistic and the mass
approaches the value of eq.~(\ref{eq:chandrasekhar}) which is the famous
{\em Chandrasekhar} mass $M_{ch}$. It represents the maximum possible
mass for a white dwarf for a purely relativistic polytropic EoS.

\subsubsection{General Relativity Corrections}
\label{GNR}

If the stars are very compact one has to take into account effects from
general relativity, like the curvature of space--time. These effects
become important when the factor ${2GM}/{c^2R}$ approaches unity.  We
then have to describe a compact star by using Einstein's equation:
\begin{eqnarray}
G_{\mu\nu} = - \frac{8 \pi G}{c^4} T_{\mu\nu}\quad .
\end{eqnarray}
For an isotropic, general relativistic, static, ideal fluid sphere in
hydrostatic equilibrium one arrives at the Tolman--Oppenheimer--Volkoff
(TOV) equation (see e.g.\ 
\cite{Weinberg72,Shapiro_book,Weber_book,Hanauske04} for detailed
derivation):
\begin{eqnarray} 
 \frac{d p}{d r} = - \frac{G \epsilon(r) {m}(r)} {c^2 r^2} 
 \left[ 1 + \frac{p(r)}{\epsilon(r)} \right]
 \left[1 + \frac{4 \pi r^3 p(r)} {{m}(r) c^2} \right]
 \left[1 - \frac{2 G {m}(r)} {c^2 r} \right]^{-1}\quad .
 \label{eq:DEpressureGR}
\end{eqnarray}
This equation is similar to the differential equation for the pressure,
eq.~(\ref{eq:structure_p}), but has three correction factors. All three
correction factors are larger than one, i.e.\ they strengthen the term
from Newtonian gravity. The TOV equation contains also the factor
${2GM}/{c^2R}$ which determines whether one has to take into account
general relativity or not. The corresponding critical radius
\begin{eqnarray}
R & = & \frac{2GM}{c^2}
\end{eqnarray}
is the so called Schwarzschild radius. For a star with a mass of 1 $
{\rm M_\odot} $, we get $R \approx 3$km. Hence effects from general
relativity will become quite important for neutron stars as they have
radii around 10 km. The corrections are smaller for white dwarfs with
radii of $R(M=1 M_{\odot}) \approx 10^4 - 10^3$ km but will be checked
numerically later. For completeness, note, that an anisotropy in the
pressure can generate an additional term to the TOV equations even for
spherical symmetry \cite{Herrera97} which we, however, do not consider
in the following.

\subsection{Calculations}

\subsubsection{Principle}

In the following, we use the structure equations (\ref{eq:structure_p})
and (\ref{eq:structure_M}) for our numerical calculations of the mass
and the radius of white dwarfs and neutron stars. We need two initial
values for the calculation, $p(0)$ and $m(0)$. From equation
(\ref{eq:ddm}) it is clear that $m(0) = 0$. We will increase $r$ in
small steps and for each step in radius the new pressure and new mass
are calculated via:
\begin{eqnarray}
p(r + \Delta r) = - \frac{G\epsilon(r)m(r)}{c^2r^2}{\Delta r}+p(r) 
\end{eqnarray}
\begin{eqnarray}
m(r + \Delta r) = \frac{4 \pi r^2 \epsilon(r)}{c^2} {\Delta r} + m(r)
\end{eqnarray}
with $\Delta r$ as a fixed constant which determines the accuracy of the
calculation. This is the so called Newton method, whereas later we also
employed a four--step Runge--Kutta method which allows a more
sophisticated numerical integration. As mentioned before in section
\ref{Structure equations} the pressure will decrease with increasing
$r$. At a certain point, the pressure $p(R)$ will become zero (or even
negative) which will terminate the integration and determines the radius
of the star $R$ and as its mass $M=m(R)$. For our computation we worked
with Fortran and Matlab.

\subsubsection{Polytropic EoS} 

First, we will use the polytropic equation of state
\begin{eqnarray}
p = K\epsilon^{\gamma}\quad .
\label{eq:polytrope}
\end{eqnarray}
In this subsection we closely follow \cite{Silbar04}. We will just
briefly discuss the method which was used in \cite{Silbar04} as well
as our results and refer the reader to the original paper for more
detailed information. The complication in solving the differential
eqs.~(\ref{eq:structure_p}) and (\ref{eq:structure_M}) is to find a way
to get for a certain pressure $p(r)$ the corresponding energy density
$\epsilon(r)$. The polytropic form of the EoS, eq.~(\ref{eq:polytrope}),
avoids this problem and being plugged in eq.~(\ref{eq:structure_p})
gives:
\begin{eqnarray}
 \frac{d{p}(r)}{d r} = 
 - \frac{R_0 p(r)^{1/\gamma} \bar{m}(r)}
 {r^2 K^1/\gamma} \, , \label{eq:DEpresDimless}
\end{eqnarray}
with:
\begin{equation}
 \bar{m}(r)= \frac{m(r)}{M_\odot}, \ \quad {R_0 =
 \frac{G {\rm M_\odot}}{c^2}} 
 \label{eq:alphadef}
\end{equation}
where ${\rm M_\odot}$ is the mass of the sun, hence $\bar{m}(r)$ is
dimensionless, and $R_0$ is half of the Schwarzschild radius and is
given in km. We also get a new expression for the differential equation
for the dimensionless mass:
\begin{equation}
 \frac{d \bar{m}(r)}{dr} = \frac{4 \pi r^2}{M_\odot c^2} \left(
 \frac{p(r)}{K}\right)^{1/\gamma}\quad , 
 \label{eq:DEcurlyMDimless}
\end{equation}
The units for energy density ${\epsilon(r)}$ and the pressure ${p(r)}$
are ${dyne/cm^2}$, the mass is expressed in ${\rm M_\odot}$ and the
radius in ${km}$. We can now calculate the mass and the radius of a
white dwarf for a given central pressure.

\begin{figure}[!ht]
{\centering \includegraphics[width=0.6\textwidth]
{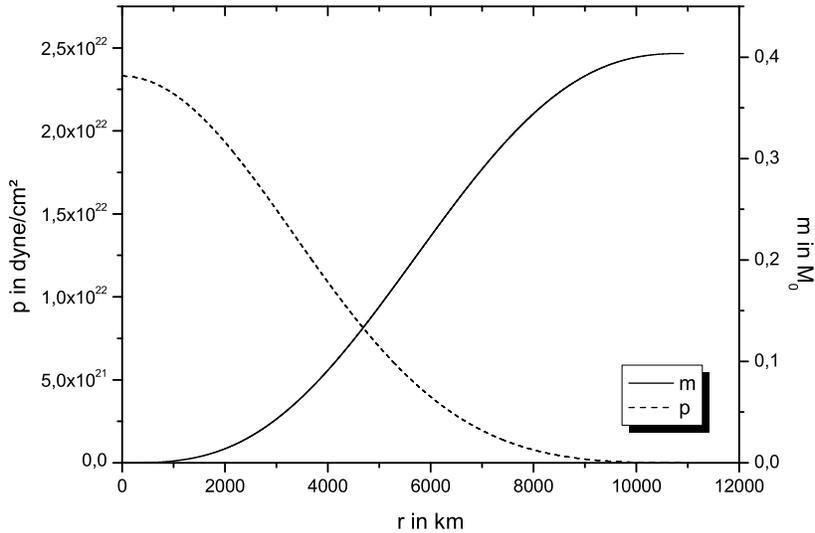} \par}
\caption{The mass $m(r)$ and the pressure $p(r)$ for a non--relativistic
  white dwarf with a central pressure of $p_0 = 2.2 \cdot 10^{22}$
  dyne/cm$^2$.}
\label{non_rel_mass}
\end{figure}

We will distinguish between the relativistic case and the
non--relativistic case by setting $\gamma = 4/3$ and $\gamma = 5/3$,
respectively. Choosing for the central pressure $p_0 = 2.33002\cdot
10^{22}{\ \rm dyne/cm^2}$, where this value satisfies the condition $k_F
\ll m_e c$ (cf.\ eq.~(\ref{eq:electronpres})), our numerical calculation
with the Newtonian Code gives us the curve shown in
Fig.~\ref{non_rel_mass} for the pressure--radius and the mass--radius
relations. The white dwarf's mass is $M = 0.40362\ {\rm M_\odot}$ and
the radius $R = 10,919$ km.

\begin{figure}[!ht]
{\centering \includegraphics[width=0.6\textwidth]
{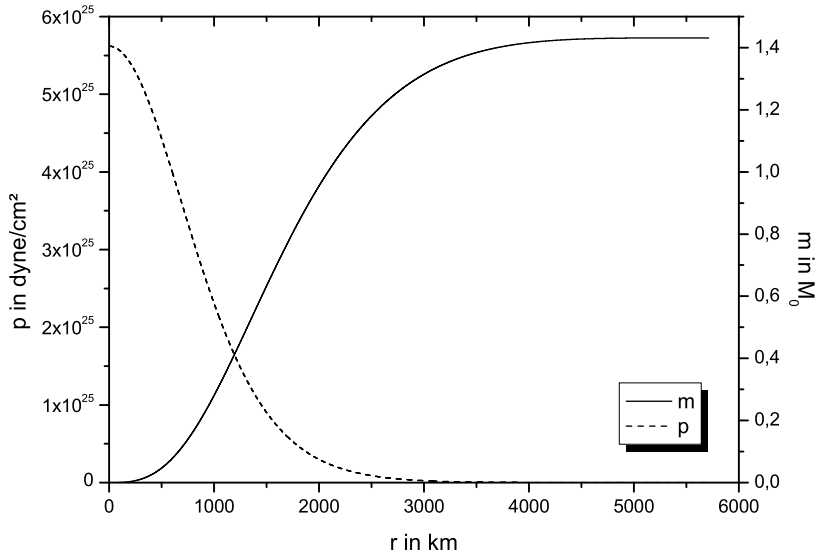} \par}
\caption{The mass $m(r)$ and the pressure $p(r)$ for a relativistic
 white dwarf with a central pressure of $p_0 = 5.62 \cdot 10^{25} {\
 \rm dyne/cm^2}$.} 
\label{rel_mass}
\end{figure}

To satisfy the condition $k_F \gg m_e c$ for a relativistic white dwarf, 
we have to choose a larger central density
compared to the non--relativistic case, $p_0 = 5.619 \cdot
10^{25}$~dyne/cm$^2$ is an appropriate value (see Fig.~\ref{rel_mass}).

There is one interesting point as far as the relativistic polytrope is
concerned. As we have seen in section~\ref{chandra} the mass limit to a
white dwarf must be 1.4312 ${\rm M_\odot}$ if we set $\eta$ equal 2 and
use a relativistic polytropic form for our equation of state. As these
are exactly the same underlying assumptions as in our numerical
computation, the maximum calculated mass must necessarily be the
Chandrasekhar mass. Indeed, our numerical calculation gives us a white
dwarf with a radius of 5710 km and a mass of 1.43145~M$_\odot$. With the
chosen high initial pressure value we hit the Chandrasekhar mass limit
very well. 

\begin{figure}[!ht]
{\centering \includegraphics[width=0.6\textwidth]
{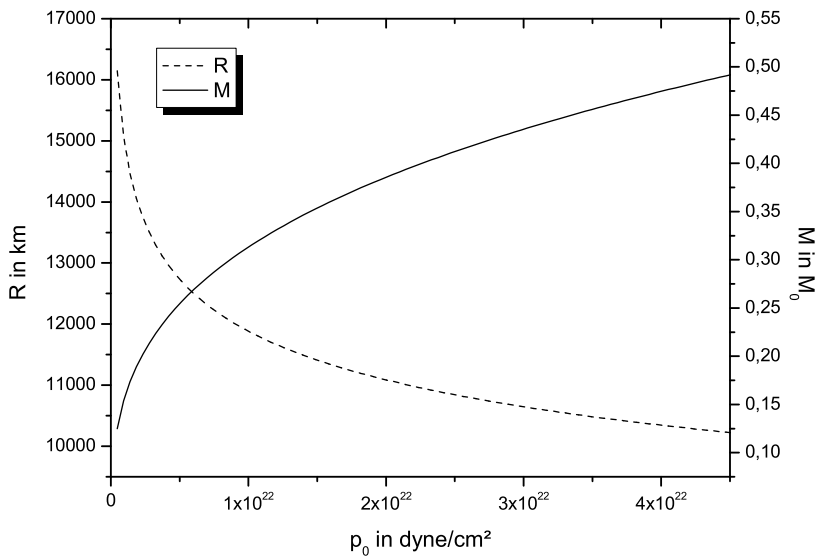} \par}
\caption{Calculated masses and radii for white dwarfs using the
 non--relativistic polytrope.} 
\label{mass_range_non_rel_pol}
\end{figure}

\begin{figure}[!ht]
{\centering \includegraphics[width=0.6\textwidth]
{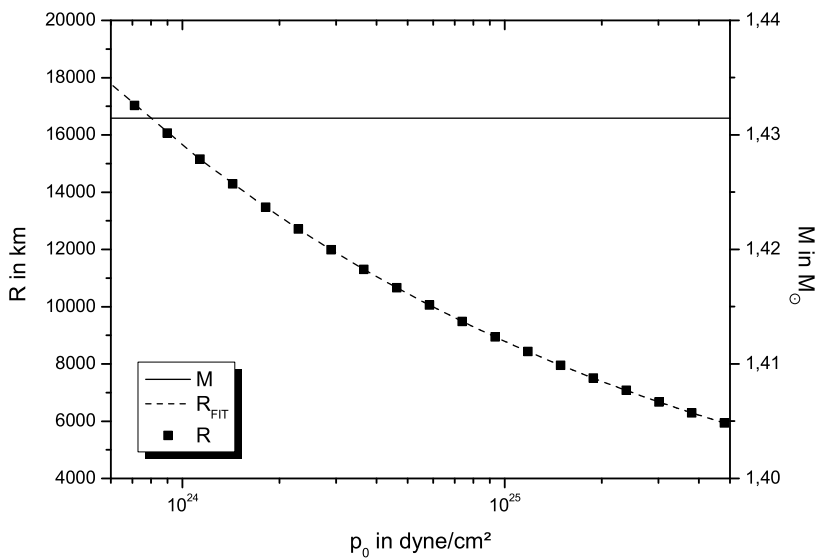} \par}
\caption{Calculated masses and radii for white dwarfs using the
 relativistic polytropic EoS.} 
\label{mass_range_rel_pol}
\end{figure}

Note, however, that according to the expression (\ref{eq:chandrasekhar})
the mass does not depend on the central pressure. To see this in more
detail we have employed different central pressures:
Fig.~\ref{mass_range_rel_pol} shows the calculated masses of white
dwarfs for initial pressures from $1 \cdot 10^{23} {\ \rm dyne/cm^2} $
to $2.5 \cdot 10^{24} {\ \rm dyne/cm^2}$. For this range of central
pressures we always compute the same mass of 1.43145 ${\rm M_\odot}$.
Hence, the numerical calculations agree with theory. As an additional
check we also plot the corresponding radii which we computed numerically
(points referred to as "R") and calculated analytically with
eq.~(\ref{eq:chandra_radius}) (referred to as "${\rm R_{FIT}}$"). The
deviations of the numerically computed radii from the analytical ones
are about 0.2\%.  Taking a look at the plot, we find that for increasing
central pressure the mass of the relativistic white dwarf stays constant
while the corresponding radius decreases. The behaviour of the mass and
radius of a non--relativistic white dwarf with smaller central pressures
is shown in Fig.~\ref{mass_range_non_rel_pol} for comparison.  In
contrast to the relativistic case the mass increases now with increasing
central pressures.

Silbar and Reddy already pointed out in \cite{Silbar04} that the two
polytropic white dwarf models examined here are not quite physical, as
the non--relativistic EoS just works for central pressures below a
certain threshold while the relativistic EoS is not applicable for small
pressure. They suggest finding an EoS which covers the whole range of
pressures by describing the energy density as a combination of a
relativistic and a non--relativistic polytrope. We decided to try another
way. On the other hand, we will employ not a guessed combination but a
consistent relation ${p(\epsilon)}$.

\subsubsection{Looking for zero--points}

We remind the reader that it is our goal to find a relation between the
energy density and the pressure which is needed for the calculation of
the mass and the radius of the star. In the following, we want to use
the full equations for a degenerate electron gas to derive the EoS
consistently for the whole pressure range considered for white dwarfs.
As we see from eqs.~(\ref{eq:electroneps}) and (\ref{eq:electronpres}),
it is impossible to write explicitly the energy density as a function of
the pressure without using certain approximations on $\epsilon$ or p
just like it was done in the previous section (i.e.\ the polytropic
approximation).

On the other hand, a numerical solution to this problem is straight
forward. As both the pressure and the energy density are functions of
the Fermi momentum, we just have to find for a given pressure the
corresponding Fermi momentum and use it in the equation for the energy
density. With this procedure we are able to calculate the energy density
$\epsilon$ for a given pressure $p$. To get the Fermi momentum we apply
a root--finding method for the equation
\begin{eqnarray}
p(k_F) - p = 0,
\end{eqnarray}
where $p$ is given and $p(k_F)$ taken from eq.~(\ref{eq:electronpres}).
As long as we have a simple expression for the derivative of pressure we
use the Newton--Raphson--Method. In other cases, when we have more
complicated equation of states, we take a simple bisection method, which
is less accurate, but easier to handle. In addition, we write the
energy density as the sum of the nucleon energy density and the electron
energy density (in the polytropic approximation the energy density of
the electrons was ignored). Employing this numerical prescription to
the solution for the EoS, we can calculate the mass/pressure and the
radius/pressure dependence, just like in the polytrope case.

\begin{figure}[!ht]
{\centering \includegraphics[width=0.6\textwidth]
{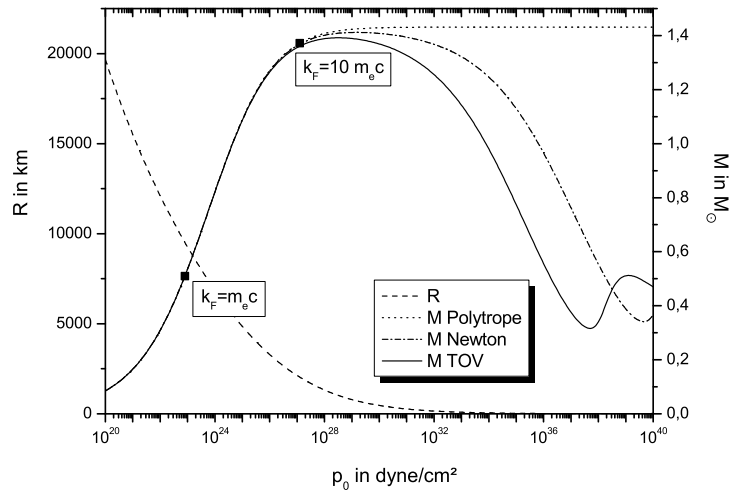} \par}
\caption{Calculated masses and radii for white dwarfs using a consistent EoS.}
\label{tov_newton}
\end{figure}

In section~\ref{GNR} we discussed the corrections from general
relativity and we will now test their influence on the masses and radii
of white dwarfs. Fig.~\ref{tov_newton} shows our calculations for three
different cases. In the case referred to as "Polytrope" the energy
density is given by just the nucleons, like in the polytropic EoS
earlier. The "Newton" mass/pressure curve is calculated with an energy
density consisting of both, the nucleon and the electron energy density
(see eq.~(\ref{eq:epsilon})), for the "TOV" curve the EoS is like the
one for "Newton" but with corrections from general relativity
(eq.~(\ref{eq:DEpressureGR})). All three curves confirm the
Chandrasekhar mass limit.

For pressures smaller than $10^{27} {\ \rm dyne/cm^2}$ the three cases
are indistinguishable while for larger pressures the curves start to
deviate significantly from each other. We have indicated two points on
the mass/pressure curves: the first one corresponds to a central
pressure where $k_F = m_e c$, i.e.\ the point where the electrons start
to become relativistic, the second point is at $k_F = 10 m_e c$ and
marks a central pressure where the electrons are highly relativistic.
At the latter point, one can already see how the three curves start to
deviate from each other. The similar behaviour for low pressure is due to
the fact, that the electron contribution to the energy density as well
as the corrections from general relativity are negligible. This is the
non--relativistic region and the behaviour of the curves is comparable to
the one of the non--relativistic polytrope ($p\propto \epsilon^{5/3}$),
the mass increases with the central pressure.

The difference of the three curves in Fig.~\ref{tov_newton} starts to
appear around $p_0 = 10^{27} {\ \rm dyne/cm^2}$ where $k_F = 10 m_e c$.
Nevertheless, in all three cases the masses stay almost constant at a
value of about $1.4\ {\rm M_\odot}$ around $p_0 = 10^{28} {\ \rm
  dyne/cm^2}$. At this pressure the relativistic electrons dominate the
mass/pressure relation like in the relativistic polytrope case
($p\propto \epsilon^{4/3}$).

We will discuss the behaviour at larger pressure starting with the case
"Polytrope". As the energy density for this model stems solely from the
nucleons, the results for mass and radius for higher pressures have to
merge into that of the relativistic polytrope. This is confirmed
by the constant mass of 1.4314 solar masses (equal to the Chandrasekhar
mass) for central densities above $p_0= 10^{31} {\ \rm dyne/cm^2}$. The
mass in the center of the white dwarfs increases with larger central
densities while the radius decreases, such that the total mass stays
constant.

For high pressures the contribution of the relativistic electrons to the
energy density becomes important which is taken into account in the
"Newton" case. The EoS for electrons only can be approximated by an
ultra--relativistic polytrope (see eq.~(\ref{eq:ultra})):
\begin{equation}
p = \frac{1}{3} \epsilon \quad .
\end{equation}
With increasing central pressure the range in radius of the white dwarf
where the electrons obey the ultra--relativistic limit enlarges.  The
deviation from the "polytrope"--curve is an effect of the growing
fraction of ultra--relativistic electrons. The masses for high pressures
are now smaller than the ones in the "Polytrope" case for the following
reason: the degeneracy pressure of electrons for the same energy density
will be lower in the ultra--relativistic polytrope compared to the
"Polytrope" case due to additional contribution of electrons to the
energy density. Hence, the star will support less mass and the mass
decreases as a function of pressure beyond the maximum mass. For the
"TOV" mass/pressure curve we take the former case adding effects from
general relativity (see eq.~(\ref{eq:DEpressureGR})). At a pressure of
$p_0 \sim 10^{28}{\rm {dyne/cm^2}}$ the mass curve of white dwarfs
exhibits a clear maximum with a maximum mass being a little bit lower
than for the Newtonian cases i.e.\ $M_{max} = 1.39\ {\rm M_\odot}$.  The
slightly reduced maximum mass is in accord with the fact that the
corrections from general relativity strengthen the gravitational force.
With increasing central pressures and smaller radii, corrections become
larger. Therefore, the mass of the white dwarf decreases with increasing
central pressures up to a pressure of about $p_0 = 2 \cdot 10^{37} {\ 
  \rm dyne/cm^2}$, here the mass starts to rise to a maximum value of
about 0.51 solar masses and then decreases again.  The general
relativistic corrections applied in the "TOV"--curve enhance the
gravitational force, so that less mass is needed to create the given
central pressure.

\begin{figure}[!ht]
{\centering \includegraphics[width=0.6\textwidth]{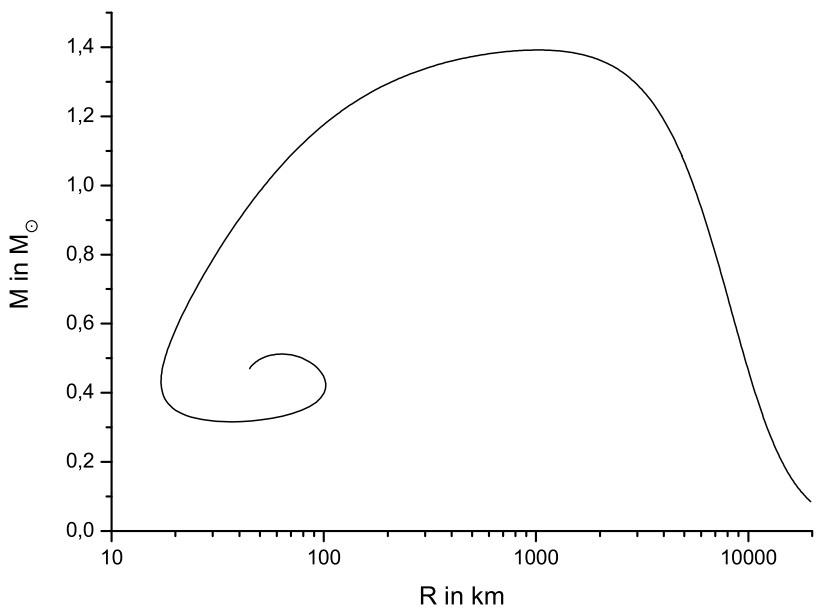} \par}
\caption{The mass--radius relation for white dwarfs with corrections for
 general relativity.} 
\label{MRwd}
\end{figure}

Fig.~\ref{MRwd} shows the calculated masses for the "TOV" case in
dependence of the radius. The central pressure increases along the curve
starting from the right to the left in the plot. The curve begins at
large radii and small masses and continues to smaller radii and larger
masses. Passing the maximum mass the curve starts to decrease with
radius. For higher central pressures the curve starts to curl. This
behaviour is typical for large central pressures, seen also in many other
mass--radius relations for different equations of state. A detailed
analysis of this curled shape can be found in \cite{Harrison65} and
\cite{Harrison_book}.
 
In the next subchapter we will discuss that white dwarfs of the "Newton"
and "TOV" case which are beyond the maximum mass limit (i.e.\ for $p \gg
p_0$) are unstable. It can be shown that polytropic equations of state
with an exponent less than $4/3$ cause an instability (see e.g.
\cite{Shapiro_book}): an increase in energy density causes an insufficient
increase in pressure to stabilise the star.

\subsubsection{Stability}

\begin{figure}[!ht]
{\centering \includegraphics[width=0.6\textwidth]
{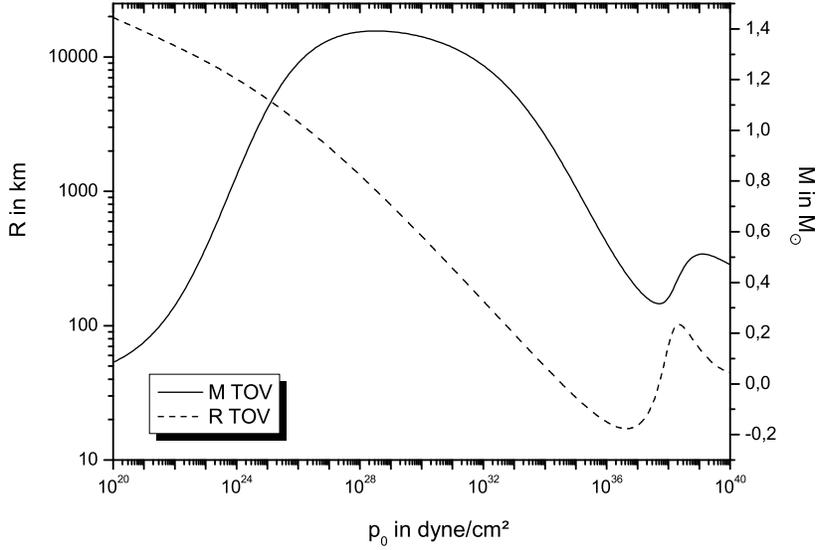} \par}
\caption{Calculated masses and radii for white dwarfs with corrections
 for general relativity} 
\label{mass_radius}
\end{figure}

First we argue that white dwarfs are unstable if $dM/dp_0 < 0$ following
\cite{Glen_book}. For our argumentation we employ the mass curve in
Fig.~\ref{mass_radius} and reason at first why white dwarfs with $p_0 \le
p_0^{max}=p_0(M=M_{max})$ are stable.

Let us assume that a white dwarf oscillates radially and therefore
increases his central pressure to $p_0^\prime = p_0 + \delta p_0$ with
$\delta p_0 > 0$. In Fig.~\ref{mass_radius} the mass curve shows white
dwarfs in hydrostatic equilibrium i.e.\ the degeneracy pressure and the
gravitational force compensate each other. If for $p_0 \le p_0^{max}$ the
central pressure is increased, the white dwarfs are only in hydrostatic
equilibrium if their masses are larger than before. Since this is not
the case, the surplus in degeneracy pressure over the gravitational
attraction will expand the star back to his former state. For pressures
$p_0^\prime = p_0 - \delta p_0$ Fig.~\ref{mass_radius} shows that the
mass $M(p_0^\prime)$ of a white dwarf in hydrostatic equilibrium is
smaller than $M(p_0)$. That means that if the star expands due to radial
oscillations the resulting lower central pressure can only be maintained
if the gravitational force, i.e.\ the mass, decreases. As the mass of
the white dwarf does not change the star has to shrink again to its
former state. Therefore, white dwarfs with $p_0 \le p_0^{max}$ and $dM/dp_0
> 0$ are stable.

Now we discuss the situation for a white dwarf with $p_0 >p_0^{max}$ and
$dM/dp_0 < 0$. If the central pressure is lowered to $p_0^\prime = p_0 -
\delta p_0$ then $M(p_0^\prime)$ should be larger than $M(p_0)$ to bring
the star in hydrostatic equilibrium (see Fig.~\ref{mass_radius}). So the
gravitational force is not large enough to stop the expansion and the
white dwarf is unstable to radial oscillations. For an increased central
pressure $p_0^\prime = p_0 + \delta p_0$ the mass curve shows that
stability is only given for a white dwarf with $p_0 \le p_0^{max}$. 
As the mass of the white dwarf is larger than the one required for
hydrostatic equilibrium, gravitation will cause the white dwarfs to
shrink and finally to collapse.

In \cite{Harrison_book}, \cite{Weber_book} as well as in \cite{Shapiro_book},
which we will follow now, a more formal approach is presented. Again,
one considers radial oscillations of the white dwarf. The radial modes
$\omega_n$ can be derived by solving a Sturm--Liouville eigenvalue
equation with the eigenvalues ${\omega_n}^2$. In doing so one gets a
range of eigenvalues with
\begin{eqnarray} 
{\omega_0}^2 < {\omega_1}^2 < {\omega_2}^2 < ... \label{omegas}
\end{eqnarray}
where $\omega_n$ are the eigenfrequencies to the radial modes and
$\omega_0$ corresponds to the fundamental radial mode. For the
instability analysis the so called "Lagrangian displacement"
{\boldmath$\xi$} (\textbf{x},t) is introduced in ref.~\cite{Shapiro_book}
which connects fluid elements in the unperturbed state to corresponding
elements in the perturbed state. The normal modes of an oscillation can
then be written as
\begin{eqnarray}
\xi^i (\textbf{x},t) = \xi^i (\textbf{x}) e^{i \omega t}\quad ,
\end{eqnarray}
where $\xi^i$ are any Lagrangian displacements. Hence, the star will be
stable for $\omega^2 \ge 0$ and unstable for $\omega^2 < 0$. This means that for stability all 
modes need to have real eigenfrequencies. 
But because of (\ref{omegas}) the stability of the star depends only on the sign of ${\omega_0}^2$.
 
The instability condition can be connected to the total Mass 
of any single star in in hydrostatic equilibrium (i.e. every star lying on our $M(R)$ curve). 
The variation $\delta M$ of the mass out of the state of equilibrium can be expressed as follows:
\begin{equation}
M = M_0 + \delta M + \delta^2 M \quad ,
\label{eq:pertuberation_energy}
\end{equation}
 
where $M_0$ is the total mass at equilibrium. 
For all stars in hydrostatic equilibrium it holds that the variation $\delta M = 0$.
It can be shown (see \cite{Shapiro_book}) that stability is given for
$\delta^2 M > 0 $ with an onset of instability at $\delta^2 M = 0 $.
For $\delta^2 M < 0 $ the star is unstable. 
In the mass/pressure relation one finds minima and maxima, where $dM/dp_0 = 0$. 
For these extrema it can be deduced that also $\delta^2 M = 0$. This means these extrema are
the critical points for stability. The latter condition relates to $\omega^2 = 0$ and therefore
signals a change in the stability of the corresponding normal mode.

In the case of multiple critical points it is also possible that an unstable 
configuration can change back to stability at the following extremum. Therefore one has to 
take into account the sign of the change in the radius R (i.e.\ $dR/dp_0 < 0$ or $dR/dp_0 > 0$) in addition. 
If an equilibrium configuration of a star is now transformed
from the low--pressure side of the critical point to the high--pressure
side, one can find the following relations between the change of the radius
$R$ with respect to the central pressure $p_0$ and the radial oscillation
modes:

\begin{itemize}
\item $dR/dp_0 < 0$: the square of the oscillation frequency of an even numbered mode changes the sign (${\omega_n}^2\rightarrow -{\omega_n}^2$)    
\item $dR/dp_0 > 0$: the square of the oscillation frequency of an odd numbered mode changes the sign.
\end{itemize}

For the "TOV" curve in Fig.~\ref{tov_newton} there are three critical
points. The first critical
point appears at the maximum around $p_0 = 10^{28} {\ \rm dyne/cm^2} $.
As can be seen in Fig.~\ref{mass_radius}  $dR/dp_0$ is negative in this critical point, therefore an
even mode is changing stability. Because of (\ref{omegas}) and the assumption that for low densities the ocsillation frequencies of all modes are real (i.e. ${\omega_n}^2 > 0$), ${\omega_0}^2$ should change its sign here.
Therefore, for central pressures between the first critical point
and the next one, the oscillation in the fundamental mode is unstable.

At $p_0 \sim 10^{37.5} {\ \rm dyne/cm^2}$, the second critical point is
reached. Here $dR/dp_0 > 0$ and ${\omega_1}^2$ becomes negative. At the
second maximum at $p_0 \sim 10^{39} {\ \rm dyne/cm^2}$ again the
oscillation in an even mode must change the sign and as ${\omega_1}^2$
is negative already, ${\omega_2}^2$ becomes negative now. This stability
analysis can also be applied to the case of the Newtonian white dwarfs
as well as to the neutron stars. As mentioned before, the sequence
can become stable again; then at the second critical point the
fundamental mode changes its sign back and a stable sequence of
compact stars starts again. This so called third family of compact stars
appears for a strong first order phase transition to e.g.\ quark matter
\cite{Gerlach68,GK2000,Schertler00}.

In summary white dwarfs become unstable as soon as their mass is above
the Chandrasekhar mass limit. This is exactly what is happening in the
case of a core collapse supernova explosion. It is quite interesting to
see that such an intuitive reasoning already indicates what is happening
in real nature.

\section{Pure Neutron Stars}
\label{ns}

\subsection{Non--relativistic polytropic case}

\begin{figure}[!ht]
{\centering \includegraphics[width=0.6\textwidth]
{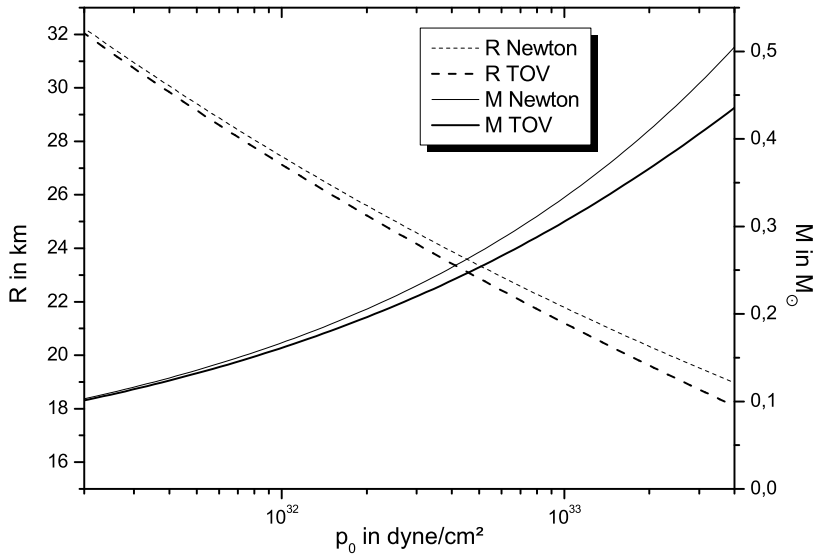}}
\caption{The mass $M$ and radius $R$ of neutron stars in the
 non--relativistic case as a function of the central pressure $p_0$. The
 results from the TOV equation are compared to the Newtonian limit.} 
\label{nrelTOV-newton}
\end{figure}

In this chapter we start looking at pure non--relativistic neutron
stars, which are described by an EoS of a Fermi gas of neutrons. In the
following we drop the subscript $F$ for labelling the Fermi momentum due
to practical reasons. With the neutron number density
\begin{equation}
 n_n=\frac{k_n^3}{3\pi^2\hbar^3}
\end{equation}
and the mass density
\begin{equation}
 \rho=n_n\cdot m_n
\end{equation}
we get in analogy to the Fermi gas for electrons (see eqs.~(\ref{eq:EF})
-- (\ref{eq:electronpres})):
\begin{eqnarray}
\epsilon(x) &=& \frac{\epsilon_0}{8}[(2x^3+x)(1+x^2)^{1/2}-\sinh^{-1}(x)]
\label{eq:e} \\ 
p(x)&=&\frac{\epsilon_0}{24}[(2x^3-3x)(1+x^2)^{1/2}+3\sinh^{-1}(x)]
\label{eq:p} 
\end{eqnarray}
with 
\begin{equation}
x=\frac{k_n}{m_nc}, \quad 
\epsilon_0=\frac{m_n^4c^5}{\pi^2\hbar^3}\quad .
\end{equation}
In the non--relativistic case eqs.~(\ref{eq:e}) and (\ref{eq:p}) simplify to:
\begin{equation}
\epsilon(x)\simeq\rho c^2=\frac{\epsilon_0}{3}x^3, \quad 
p(x)\simeq\frac{\epsilon_0}{15}x^5.
\end{equation}
Hence, the EoS can be described by a polytrope:
\begin{equation}
p(\epsilon)=K_{nrel}\epsilon^{5/3}
\end{equation}
with
\begin{equation}
K_{nrel} = \frac{\hbar^2}{15 \pi^2 m_n}\left( \frac{3 \pi^2}{m_n c^2}
\right)^{5/3} = 6.428 \cdot 10^{-26} \ \frac{{\rm cm}^2}{{\rm
    erg}^{2/3}}\quad . 
\end{equation}
We start our numerical calculation with our Newtonian code and the
results are plotted in Fig.~\ref{nrelTOV-newton}. We arrive at neutron
star masses of around 0.5 M$_\odot$ and radii of 20 km to 30 km. The
general relativistic effects for neutron stars should be larger than for
the white dwarfs. In Fig.~\ref{nrelTOV-newton} we compare the Newtonian
with the TOV--code results. We see that the general relativistic
corrections are at the order of a few percent and become more and more
pronounced for more massive neutron stars.

\subsection{Ultra--relativistic polytropic case}

For the ultra--relativistic limit, $x \gg 1$, one finds for the energy
density and for the pressure of nucleons:
\begin{equation}
\epsilon(x)=\frac{\epsilon_0}{4}x^4 \quad , \qquad
p(x)=\frac{\epsilon_0}{12}x^4
\Rightarrow \quad p=\epsilon/3 
\end{equation}
which is similar to the case of ultra--relativistic electrons
discussed before. However, the numerical integration of the TOV
equations fails here: The condition for terminating the integration ($p
\le 0$) can not be reached. The pressure converges monotonically as a
function of the radius against zero, so that the radius of the star
approaches infinity. The mass does not go to infinity because $dm/dr$
also reaches zero asymptotically. The problem with the relativistic case
is, that its solution has some principal inconsistencies. Running from
the initial pressure $p_0$ to zero through the whole star, the pressure
has to pass the region where neutrons become non--relativistic,
necessarily. The pure relativistic approximation is not valid for the
whole density range of a compact star.

\subsection{Full relativistic case}

\begin{figure}[!ht]
{\centering \includegraphics[width=0.6\textwidth]
{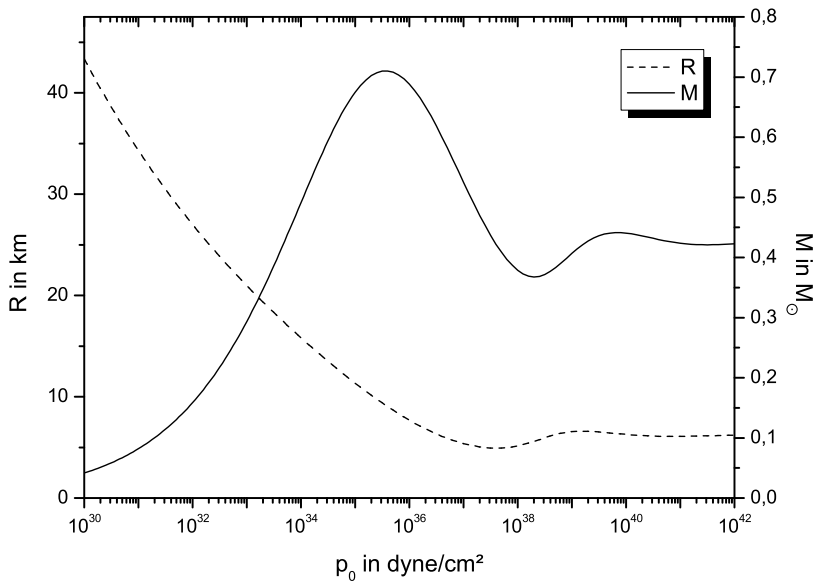}}
\caption{Mass $M$ and radius $R$ as a function of the central pressure
 $p_0$ for pure neutron stars.} 
\label{arbiRMp0}
\end{figure}

It is essentially to find an equation of state which is valid over the
whole range of pressure and energy density. In other words:
Eqs.~(\ref{eq:e}) and (\ref{eq:p}) should be used for the calculation.
Again we use a root--finding routine, as we have no explicit equation of
state of the form $p(\epsilon)$ or $\epsilon(p)$ any more. The results
for $R(p_0)$ and $M(p_0)$ are shown in Fig.~\ref{arbiRMp0}.

\begin{figure}[!ht]
{\centering \includegraphics[width=0.6\textwidth]
{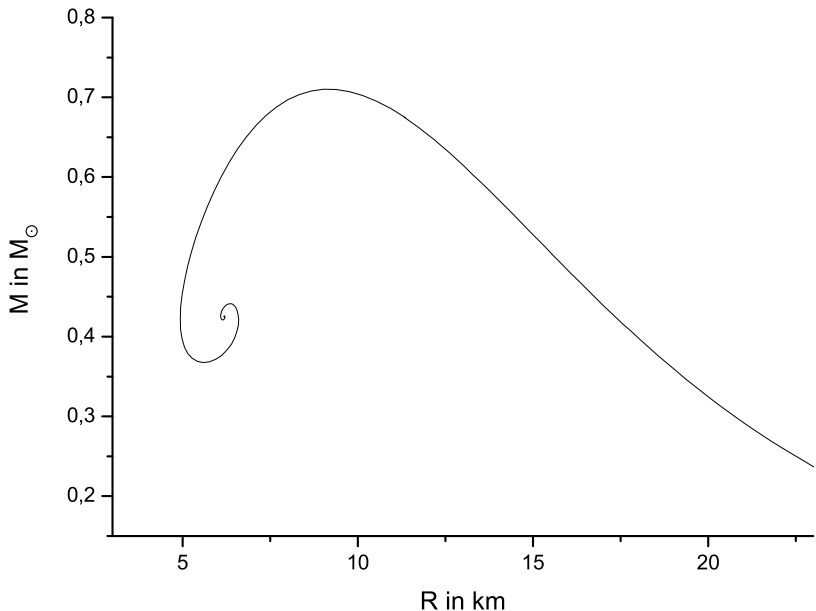}}
\caption{The mass $M$ for pure neutron stars as a function of radius $R$. The
  maximum mass is $M=0.712$ M$_\odot$ at a radius of $R=9.14$ km.}
\label{arbiRM}
\end{figure}

The mass--radius relation $M(R)$ is plotted in Fig.~\ref{arbiRM}, where
the maximum mass can be easily recognised. The heaviest stable pure
neutron star has a central pressure of $p_0=3.5\cdot10^{35}$
dyne/cm$^2$, a mass of $M=0.712$ M$_\odot$ and a radius of $R=9.14$ km
as first calculated by Oppenheimer and Volkoff in 1939 \cite{OV39} (they
arrived at a maximum mass of $M=0.71$ M$_\odot$ and a radius of $R=9.5$
km for comparison).

\section{Neutron Stars with Protons and Electrons}

\begin{figure}[!ht]
{\centering \includegraphics[width=0.6\textwidth]
{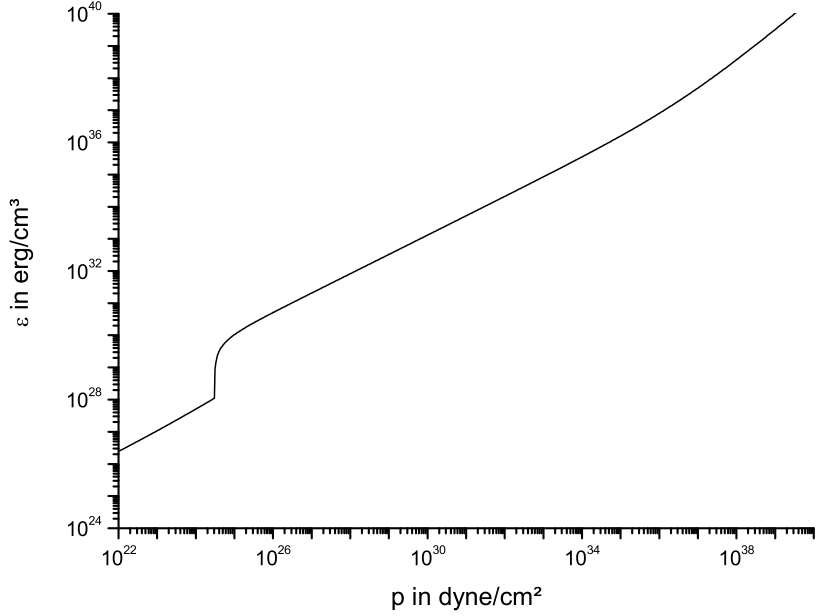}}
\caption{The energy density $\epsilon$ as a function of the pressure $p$
 for neutron stars including protons and electrons. At $p=3\cdot10^{24}
 {\ \rm dyne/cm^2}$ neutrons begin to appear.} 
\label{ep}
\end{figure}

Because the neutron is unstable, a neutron star will not consist of
neutrons only, but in addition a finite fraction of protons and
electrons will appear in dense matter. The electrons and protons are
produced by $\beta$--decay:
\begin{equation}
n\longrightarrow p+e^-+\bar\nu_e.
\end{equation}
There exist reverse reactions also, so that an electron and a proton
scatter into a neutron and a neutrino:
\begin{equation}
p+e^-\longrightarrow n+\nu_e.
\end{equation}
As all reactions are in chemical equilibrium, for a cold neutron star,
the following relation for the chemical potentials holds:
\begin{equation}
\mu_n=\mu_p+\mu_e. \label{eq:chempot}
\end{equation}
Neutrinos do not have a finite chemical potential ($\mu_{\nu}=0$), as
the neutrinos escape from the cold star without interactions.
Due to stability reasons (see e.g.\ \cite{Glen_book}), it is assumed that
the whole neutron star is electrically neutral, hence:
\begin{equation}
n_p=n_e\Leftrightarrow k_p=k_e \label{eq:charge}
\end{equation}
and the Fermi momenta of protons and electrons must be equal. With the
definition of the chemical potential for an ideal Fermi gas with zero 
temperature,
\begin{equation}
 \mu_i(k_i) = \frac{d \epsilon_i}{d n_i} = 
 (k_i^2c^2 + m_i^2c^4)^{1/2} , \quad i = n, p, e,
\end{equation}
and using eqs.~(\ref{eq:chempot}) and (\ref{eq:charge}), one arrives at
the following relation:
\begin{equation}
 (k_n^2c^2 + m_n^2c^4)^{1/2} - (k_p^2c^2 + m_p^2c^4)^{1/2} -
 (k_p^2c^2 + m_e^2c^4)^{1/2} = 0. 
\end{equation}
This equation can be solved for the Fermi momenta of protons $k_p$ as a
function of the Fermi momenta of neutrons $k_n$:
\begin{equation}
 k_p(k_n) = \frac{[(k_n^2c^2 + m_n^2c^4 - m_e^2c^4)^2 
 - 2 m_p^2c^4 (k_n^2c^2 + m_n^2c^4 + m_e^2c^4) + m_p^4c^8]^{1/2}}
 {2c (k_n^2c^2 + m_n^2c^4)^{1/2}}\quad . \label{eq:kpkn}
\end{equation}
The total energy and the pressure are now the sum of the energy and
pressure of electrons, protons and neutrons, respectively:
\begin{equation}
 \epsilon_{tot} = \sum_{i=n,p,e} \epsilon_i \, , \quad p_{tot} =
 \sum_{i=n,p,e} p_i, 
\end{equation}
where
\begin{eqnarray}
\epsilon_i(k_i) & = & \frac{8\pi}{(2\pi\hbar)^3}\int_0^{k_i} (k^2c^2 +
m_i^2c^4)^{1/2} k^2 dk \\ 
 p_i(k_i) & = & \frac{1}{3}\frac{8\pi}{(2\pi\hbar)^3}\int_0^{k_i}
 (k^2c^2 + m_i^2c^4)^{-1/2} k^4 dk. 
\end{eqnarray}
For the case when there are no neutrons present, i.e.\ $k_n=0$,
eq.~(\ref{eq:kpkn}) gives:
\begin{equation}
 k_p(0) = \frac{[(m_n^2c^4 - m_e^2c^4)^2 
 - 2 m_p^2c^4 (m_n^2c^4 + m_e^2c^4) + m_p^4c^8]^{1/2}}
 {2 m_nc^3}=1.264\cdot10^{-3}m_nc.
\end{equation}
This means, that eq.~(\ref{eq:kpkn}) can only be used, if
$k_p>1.264\cdot10^{-3}m_nc$, which is fulfilled for $p>p_{\rm
  crit}=3.038\cdot10^{24} {\ \rm dyne/cm^2}$. Below this pressure, no
neutrons are present any more, so the "neutron star" material consists
of protons and electrons only. Fig.~\ref{ep} shows the equation of state
for a free gas of protons, electrons and neutrons in
$\beta$--equilibrium in the form $\epsilon(p)$. Around the critical
pressure, a notable kink appears in the equation of state, which can be
identified with the strong onset of neutrons appearing in matter with
their corresponding additional contribution to the energy density. (One
can show that $\left.\frac{d^2\epsilon}{dp^2}\right|_{p_{\rm
    crit}}=\infty$, but $\left.\frac{d\epsilon}{dp}\right|_{p_{crit}}$
is continuous and thus the physical situation is not that of a true
thermodynamic phase transition.)

\begin{figure}[!ht]
{\centering \includegraphics[width=0.6\textwidth]
{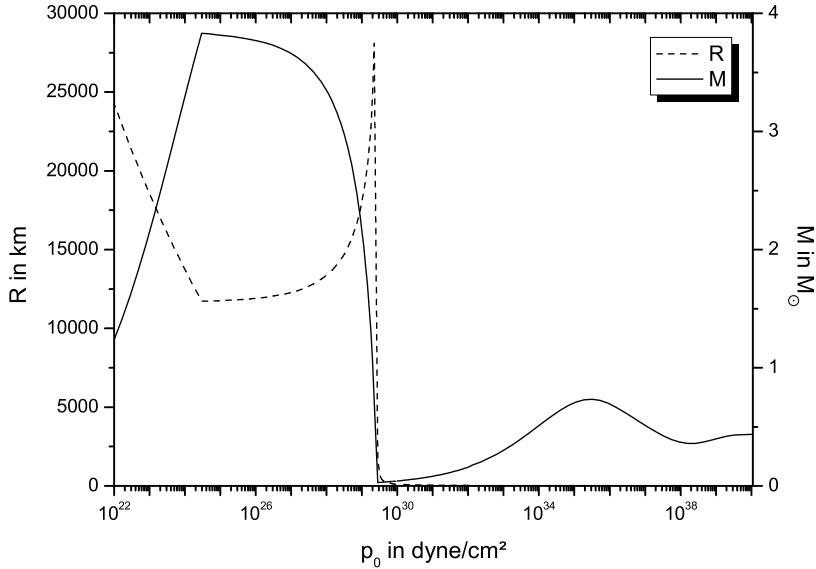}}
\caption{Mass $M$ and radius $R$ as a function of the central pressure
 $p_0$ for neutron stars including protons and electrons.} 
\label{elecpro}
\end{figure}

\begin{figure}[!ht]
{\centering \includegraphics[width=0.6\textwidth]
{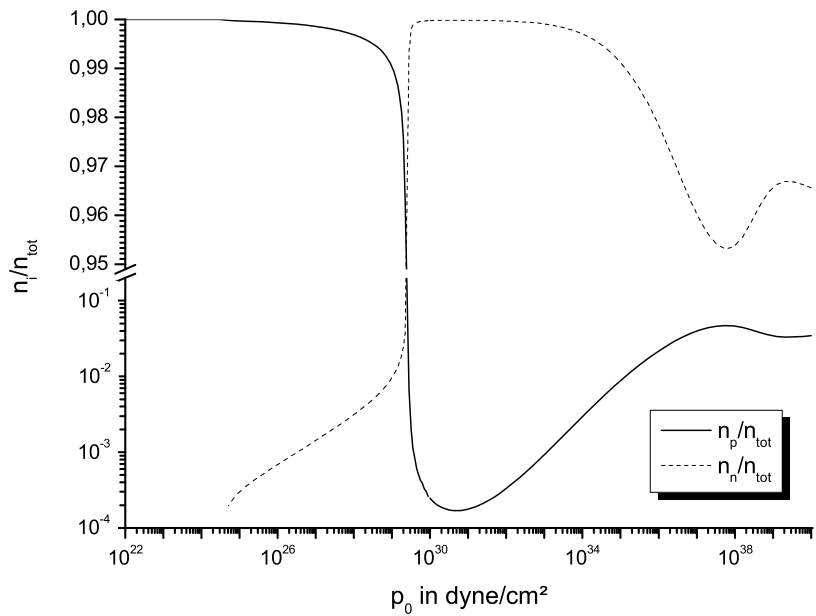}}
\caption{The proton to nucleon fraction $n_p/n_{tot}$ and the neutron to
 nucleon fraction $n_n/n_{tot}$, integrated over every single star,
 versus the central pressure $p_0$.} 
\label{npntot}
\end{figure}

\begin{figure}[!ht]
{\centering \includegraphics[width=0.6\textwidth]
{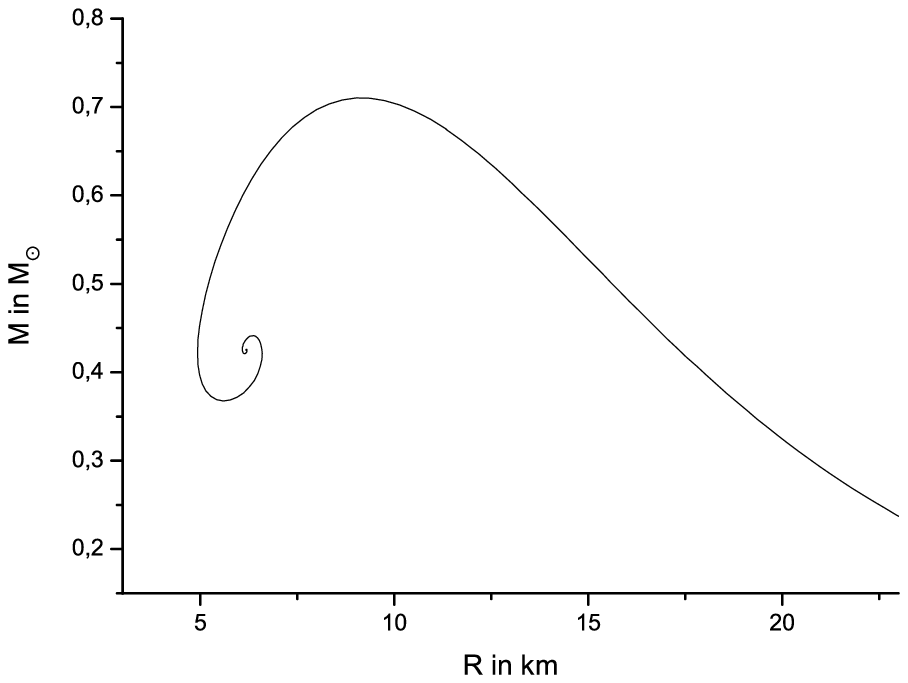}}
\caption{The Mass $M$ as a function of the radius $R$ for neutron stars
 including protons and electrons. The maximum mass star has a mass of
 $M=0.699 {\ \rm M_{\odot}}$, the corresponding radius is $R=9.24$ km.}
\label{elecproMR}
\end{figure}

Using the new equation of state, we start our TOV calculation and
compute masses and radii which are depicted in Fig.~\ref{elecpro} and
Fig.~\ref{elecproMR}, respectively. The result seems to be surprising at
a first glance but can be explained by looking at the proton to nucleon
fraction integrated over every single star as displayed in
Fig.~\ref{npntot}.

Below the critical pressure of $p=3.038\cdot10^{24} {\ \rm dyne/cm^2}$,
the neutron star consists of protons and electrons only by definition.
Above this pressure, neutrons appear, resulting in a sharp bend of
$M(p_0)$ and $R(p_0)$. Between the critical pressure and $p\simeq10^{29}
{\ \rm dyne/cm^2}$, the stars have a tiny neutron fraction in the core,
which increases with the pressure. The increasing neutron fraction is
responsible for the decreasing masses. On the microscopic scale the
relativistic electrons combine with the protons to form neutrons,
thereby reducing the electron degeneracy pressure. For the same energy
density the pressure will be lower, if neutron production is included,
so only a smaller mass is supported by the star. At $p\simeq10^{30} {\ 
  \rm dyne/cm^2}$ the proton fraction drops to $n_p/n_{tot}\leq10^{-3}$
and the star becomes an almost pure neutron star (hence the name). The
proton fraction is so small, that it can be safely neglected. As
a consequence the stellar objects become more compact, with a smaller
radius and mass. For larger pressures ($p>10^{30} {\ \rm
  dyne/cm^2}$), the mass increases again with rising pressure. The
neutrons start to become degenerate, the star is stable because of the
neutron degeneracy pressure.

\begin{figure}[!ht]
{\centering \includegraphics[width=0.6\textwidth]
{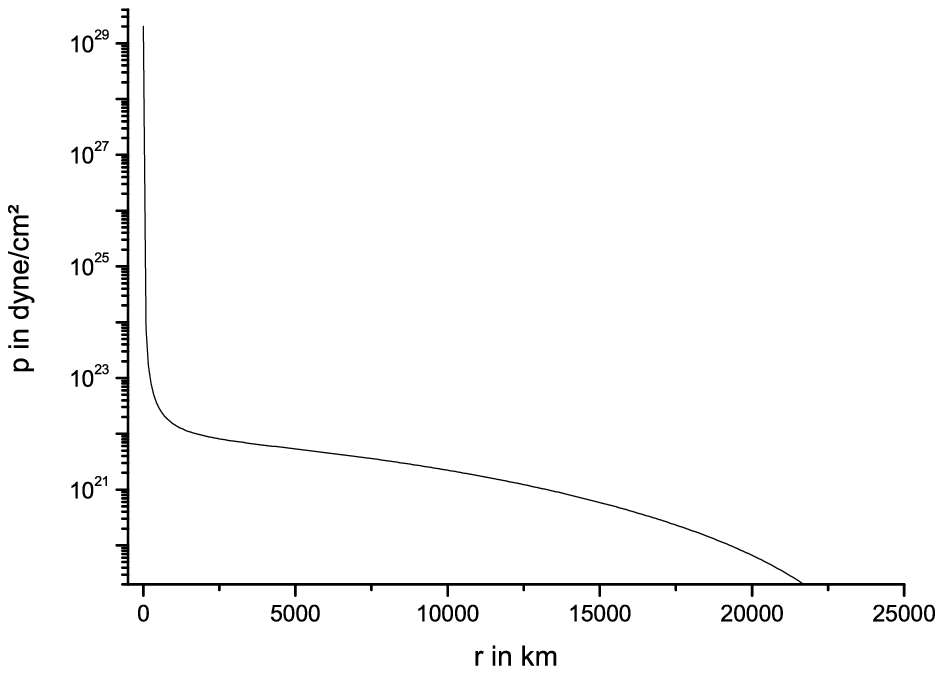}}
\caption{The pressure $p$ along the radius $r$ for a single neutron star
 including protons and electrons at $p_0=2\cdot10^{29}{\ \rm
 dyne/cm^2}$.} 
\label{mp2e29}
\end{figure} 

So the behaviour of $M(p_0)$ is understood through the change of the
proton fraction $n_p/n_{tot}(p_0)$.  But what about the unusual peak of
$R(p_0)$? Similar to the ultra--relativistic description of a pure
neutron star, the situation at $p_0\simeq10^{29} {\ \rm dyne/cm^2}$ is,
that the pressure approaches zero very slowly, but never reaches it or
falls below.  Fig.~\ref{mp2e29} shows a star with $p_0=2\cdot10^{29}{\ 
  \rm dyne/cm^2}$, where this behaviour can be clearly seen. So it is
necessary to make a cut--off (for this case at $p=10^{19} {\ \rm
  dyne/cm^2}$), so that the integration terminates as soon as this
pressure is reached.  During the numerical computation the mass
converges to $M$, while the radius grows bigger and bigger. This neutron
star consists of a dense central core and a huge low--density crust.
The protons and electrons are responsible for these huge radii: the
energy density for the same pressure is much lower, if no neutrons are
present (see Fig.~\ref{ep}). Because $dp/dr$
(eq.~(\ref{eq:DEpressureGR})) is a function of the energy density
$\epsilon$, the pressure $p(r)$ inside a star will decrease much slower,
if there are only protons and electrons left. For more massive stars
this crust disappears and the radius is determined by the neutron core.
After all, such a huge crust does not seem to be very realistic for neutron
stars.

Note that for the most compact neutron stars (i.e.\ for $p_0>10^{34} {\ 
  \rm dyne/cm^2}$) protons appear again. The proton fraction approaches
a certain constant limit at high densities. Using eq.~(\ref{eq:kpkn}) in
the ultra--relativistic limit, $k_nc \gg m_nc^2$, results in:
\begin{equation}
 k_p(k_n)\approx1/2 k_n \Longrightarrow n_p=1/8 n_n
\end{equation}
\begin{equation}
 \Longrightarrow \frac{n_p}{n_{tot}}=\frac{n_p}{n_p+n_n}=\frac{1}{9}\quad .
\end{equation}
The proton fraction for large $p_0$ in Fig.~\ref{npntot} is of course
smaller than 1/9, as only the central core obeys the relativistic limit,
and no protons are present for the regions in the star with smaller
pressure. The maximum mass star has a proton fraction of
$n_p/n_{tot}\simeq0.02$.

Another important point is, that all stars in the region of $10^{24} {\ 
  \rm dyne/cm^2}<p_0<10^{29} {\ \rm dyne/cm^2}$ are unstable, which can
be seen from the $M$ versus $p_0$ plot in Fig.~\ref{elecpro}. Hence,
there are only two different kinds of stable objects: first the
"hydrogen white dwarfs", like stars consisting of protons and electrons,
up to $M=4{\ \rm M_\odot}$ and radii of about $10^4$ km, then the almost
pure neutron stars with a maximum mass of $M=0.699{\ \rm M_\odot}$ and a
radius $R=9.24$ km at a central pressure $p_0=3.412\cdot10^{35} {\ \rm
  dyne/cm^2}$. Compared to $M=0.712 {\ \rm M_\odot}$ and $R=9.14$ km for
the pure neutron star of the last section, the differences are quite
small. The minimum neutron star mass is found to be about $0.03 M_\odot$
with a radius of around 1000 km; the radius rapidly changes for this
mass range and quickly shrinks to values of 200--300 km for slightly
larger masses. We note that these values for the minimum mass of neutron
stars are not far from the result of $M_{\rm min} = 0.09 M_\odot$ found
in more elaborate calculations (see e.g. \cite{Haensel02}).

\section{Models for the nuclear interactions}

\subsection{Empirical interaction}
\label{empirical_interactions}

In this and the following section we use MeV and fm for the energy and
distance units. Additionally $c$ is set to 1 and $\hbar c=197.3$ MeV
fm. 

Following \cite{Silbar04}, we introduce an empirical interaction for
symmetric matter ($n_n=n_p$) of the form (see also \cite{Prakash88}):
\begin{equation}
 \frac{\epsilon(n)}{n} = m_N + 
 \left<E_0\right> u^{2/3} + 
 \frac{A}{2} u + \frac{B}{\sigma + 1} u^\sigma, \quad
 u=n/n_0 \label{en}\quad . 
\end{equation}
The dimensionless quantity $\sigma$, and the quantities $A$ and $B$ with
dimension of MeV, are the fit parameters of the model. In fact, only the
third and the fourth term of eq.~(\ref{en}) describe interaction terms,
$\left<E_0\right>$ is the average kinetic energy per nucleon of
symmetric matter in its ground state: 
\begin{equation}
 \left<E_0\right> = \frac{3}{5} \frac{k_0^2}{2
 m_N}=\frac{3}{5}\frac{1}{2m_N}\left(\frac{3\pi^2\hbar^3n_0}{2}\right)^{2/3}\quad . 
\label{e0}
\end{equation}
The factor two in the denominator of the parenthesis comes in as an
additional degeneration factor for the isospin (compare to
eq.~(\ref{eq:nelectrons})).  

In the model the following values should be reproduced by the ansatz of
eq.~(\ref{en}):
\begin{itemize}
\item the equilibrium number density $n_0=0.16 {\ \rm nucleons/fm^3}
  \Longrightarrow k_0=263 {\ \rm MeV}$
\item the average binding energy per nucleon $BE = \left.(E/A - m_N)
  \right|_{n=n_0}= \left.(\epsilon(n)/n-m_N\right)|_{n=n_0}=-16 {\ \rm
    MeV}$
\item the nuclear compressibility in nuclear matter $K(n_0) = \left.9
    \frac{d p(n)}{d n}\right|_{n=n_0}=400 {\ \rm MeV}$
\end{itemize}
After some algebra one gets the following parameters:
\begin{equation}
 A = -118.2 {\ \rm MeV}, \quad B = 65.39 {\ \rm MeV}, \quad 
 \sigma = 2.112, \quad \left<E_0\right>=22.1 {\ \rm MeV}.
\end{equation}
For known $\epsilon(n)$ it is possible to determine $p(n)$ via:
\begin{equation}
 p(n) = n^2 \frac{d}{d n} \left( \frac{\epsilon}{n} \right) =
 n_0 \left[ \frac{2}{3}\left<E_0\right> u^{5/3} + 
 \frac{A}{2} u^2 + 
 \frac{B\sigma}{\sigma+1} u^{\sigma+1} \right] \label{pn}\quad .
\end{equation}
We proceed now to extrapolate to asymmetric nuclear matter.

First, the neutron to proton ratio is expressed through a parameter $\alpha$:
\begin{equation}
 \alpha = \frac{n_n - n_p}{n} = \frac{N-Z}{A}\quad . 
\end{equation}
Now the kinetic energy is given by the sum of the kinetic energies
of protons and neutrons:
\begin{eqnarray}
 \epsilon_{KE}(n,\alpha) &=& \frac{3}{5} 
 \frac{k_n^2}{2 m_N}\, n_n + 
 \frac{3}{5} \frac{k_p^2}{2 m_N}\, n_p \\
 &=& n \left<E_F\right> \frac{1}{2} \left[ 
 \left(1+\alpha\right)^{5/3} + 
 \left(1-\alpha\right)^{5/3} \right], \label{ekin} \\
 \left<E_F\right> &=&
 \frac{3}{5}\frac{1}{2m_N}\left(\frac{3\pi^2\hbar^3n}{2}\right)^{2/3}\quad . 
\end{eqnarray}
The difference in the kinetic energy density relative to the symmetric
case can be expressed as a function of $n$ and $\alpha$: 
\begin{eqnarray}
 \Delta\epsilon_{KE}(n,\alpha) &=& \epsilon_{KE}(n,\alpha) - 
 \epsilon_{KE}(n,0) \nonumber \\
 &=& n \left<E_F\right> \left\{ \frac{1}{2} \left[ 
 \left(1+\alpha\right)^{5/3} + 
 \left(1-\alpha\right)^{5/3} \right]
 - 1 \right\}\quad . 
\end{eqnarray}
This gives the following Taylor--Approximation:
\begin{equation}
 \Delta\epsilon_{KE}(n,\alpha) = n \left<E_F\right> \frac{5}{9} 
 \alpha^2 \left( 1 + \frac{\alpha^2}{27} + \cdots \right).
\end{equation}
This motivates that it is sufficient to keep corrections of order $\alpha^2$:
\begin{equation}
 \frac{E(n,\alpha)}{A} = \frac{E(n,0)}{A} + \alpha^2 S(n). \label{eemp}
\end{equation}
The function $S(u)$, which describes the symmetry energy, is assumed to
have the following form: 
\begin{equation}
 S(u) = (2^{2/3} - 1) \left<E_0\right> 
 \left( u^{2/3} - F(u) \right) + S_0 F(u). \label{su}
\end{equation}
$S_0$ is an input parameter (the asymmetry energy), which describes the
energy difference between pure neutron matter and normal symmetric
nuclear matter at ground state density $n_0$.  In the following
calculations, $S_0$ is set to 30 MeV. The only restrictions to the
function $F(u)$ are $F(1)=1$, and $F(0)=0$, in order to satisfy
$S(1)=S_0$, and $S(0)=0$, respectively. Besides this, $F(u)$ can be
chosen freely. We make the straight forward choice,
\begin{equation}
F(u)=u.
\end{equation}
The factor $(2^{2/3} - 1) \left<E_0\right>$ can be explained as follows:
For pure neutron matter, $\alpha=1$, the kinetic energy is not given by
eq.~(\ref{e0}) any more, the factor 2 in the denominator of the
parenthesis has to drop out. The factor $(2^{2/3} - 1) \left<E_0\right>$
takes this into account: 
\begin{eqnarray}
\frac{E(n,1)}{A} &=& m_N + 
 \left<E_0\right> u^{2/3} + 
 \frac{A}{2} u + \frac{B}{\sigma + 1} u^\sigma + 
 (2^{2/3} - 1) \left<E_0\right> 
 \left( u^{2/3} - F(u) \right) + S_0 F(u) \nonumber \\
 &=& m_N + \left<E_0\right> u^{2/3} + (2^{2/3} - 1)
 \left<E_0\right> u^{2/3} + \frac{A}{2} u +
 \frac{B}{\sigma + 1} u^\sigma +
 \left(S_0-(2^{2/3} - 1) \left<E_0\right> \right)F(u)
 \nonumber \\ 
 &=& m_N + 2^{2/3}\left<E_0\right> u^{2/3} + \frac{A}{2}
 u + \frac{B}{\sigma + 1} u^\sigma+\left(S_0-(2^{2/3} -
 1) \left<E_0\right> \right)
 F(u) 
\end{eqnarray} 
Thus the ansatz of eq.~(\ref{su}) implies the correct expression for
the kinetic energy of pure neutron matter. Also the different
contributions to the potential energy are separated from each other. 
 
With $\epsilon(n,\alpha)=nE(n,\alpha)/A$ it is straightforward to
calculate $p(n,\alpha)$:
\begin{eqnarray}
 p(n,\alpha) &=& n^2
 \frac{d}{dn}\left(\frac{\epsilon(n,\alpha)}{n}\right) \nonumber \\ 
 &=& p(n,0) + n_0 \alpha^2 \left[
 (2^{2/3}-1)\left<E_0\right>
 \left(2/3 u^{5/3} - u^2\right) +
 S_0 u^2 \right]. 
\end{eqnarray}
As seen in the previous section, the proton fraction for stable compact
stars is small. Hence, the parameter $\alpha$ is set to be
equal 1 and a pure neutron star is considered in the following.
 
\begin{figure}[!ht]
{\centering \includegraphics[width=0.6\textwidth]{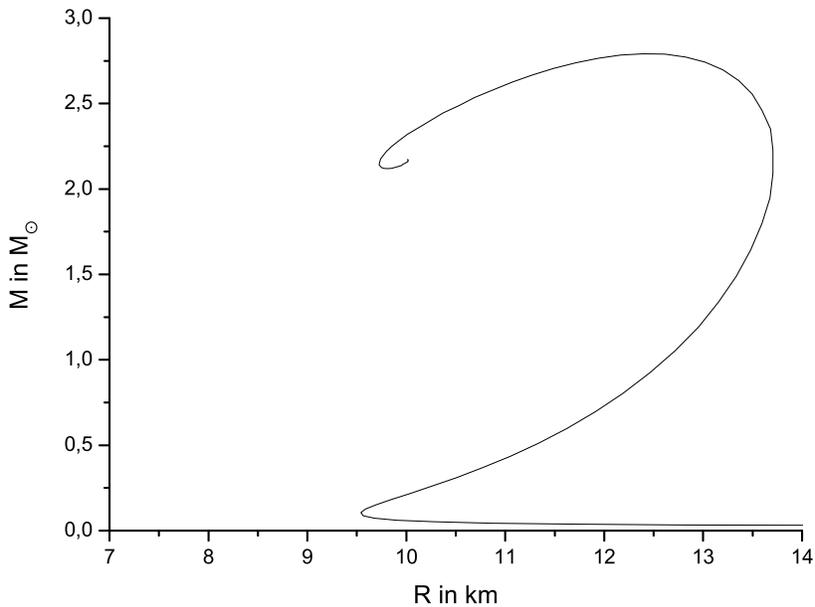}}
\caption{The mass $M$ as a function of the radius $R$ for the empirical
  nucleon--nucleon interaction. The maximum mass star has a mass of
  $M=2.792 {\ \rm M_{\odot}}$, the corresponding radius is $R=12.46$
  km.}
\label{mrempi}
\end{figure}

With $p(n)$ and $\epsilon(n)$ given, one can start again the numerical
computation. The results are shown in Fig.~\ref{mrempi}. The maximum
mass star has a mass of $M=2.792 {\ \rm M_{\odot}}$ and a radius of
$R=12.46$ km. The neutron star is much more massive, if nucleon--nucleon
interactions are included. As the equation of state is much harder,
which means that the pressure increases faster with the energy density,
the star can support a much larger mass.

\begin{figure}[!ht]
{\centering \includegraphics[width=0.6\textwidth]{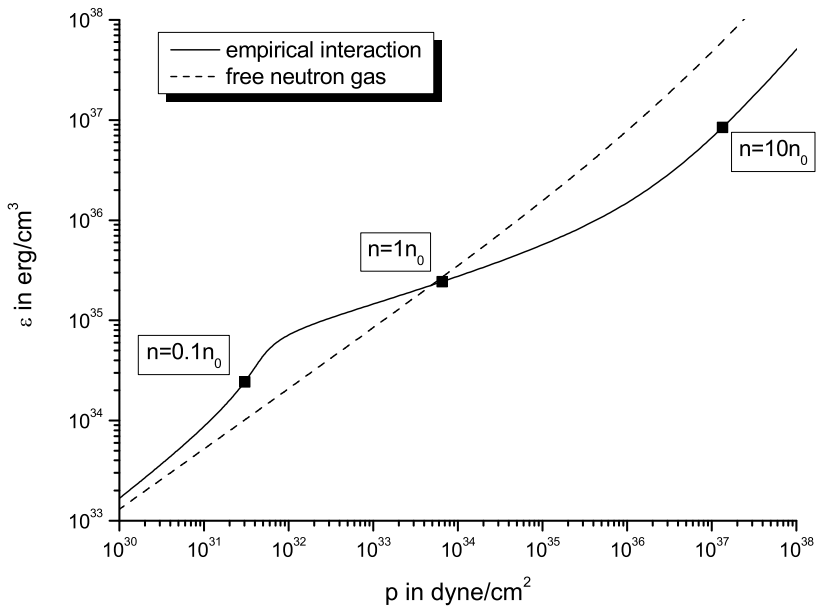}}
\caption{The equation of state for the empirical nucleon--nucleon
  interaction compared to the one of a free neutron gas.}
\label{eosempi}
\end{figure}

Another interesting fact is the behaviour of the mass for decreasing
central pressure. In a certain region the mass--radius curve drops to
smaller radii as the pressure and the mass decrease, until a radius of
$R\simeq9.5$ km and a mass of $M\simeq0.1{\ \rm M_{\odot}}$ is reached
at a central pressure of $p\simeq 1.7\cdot10^{33}{\ \rm dyne/cm^2}$.
Passing this point, the radius starts to increase again. This special
shape is due to the equation of state employed here. In Fig.~\ref{eosempi}
one sees that the EoS differs from the ones before: There is a smooth
bump at $p\simeq 4\cdot10^{31}{\ \rm dyne/cm^2}$. For pressures below
this value an increase in the energy density leads to a very small
increase in pressure, which is due to the attractive part of the
interaction for matter close to the ground state density. In this
pressure region a higher central pressure can only be generated by a
smaller radius at a nearly unchanged mass. The slope of the EoS for
pressures larger than the value of $p\simeq 4\cdot10^{31}{\ \rm
  dyne/cm^2}$ changes noticeable, so that also the mass--radius relation
for the corresponding neutron stars is affected. For larger pressures
the pressure increases much faster with the energy density, which
results in the abrupt rise of the masses (see Fig.~\ref{mrempi}). The
repulsive force of the interaction becomes dominant for nuclear
densities above ground state density. Now an increase in energy
density leads to a strong increase in pressure also and a larger mass is
supported by the star.

In fact, the EoS used here is not valid for small pressures and energy
densities. Below nuclear matter density $n_0$ it is not realistic to use
a uniform matter distribution anymore. The formation of a lattice of
nuclei and the occurance of certain possible mixed phases become
important in the outer layers, the so called crust of neutron stars. For
pressures below $10^{33}{\ \rm dyne/cm^2}$ one has to use a special
equation of state to describe the matter in the crust. Taking into
account a realistic EoS for the crust would wash out the bump seen in
Fig.~\ref{eosempi} and also the characteristic features in the
mass--radius relation described above. A good introduction to neutrons
star crusts can be found in e.g.\ \cite{BPS} and \cite{Haensel01}.

\subsection{Skyrme parameterised interaction}

In the following, the properties of nuclear matter are considered by
applying the Hartree--Fock--method on a phenomenological nucleon--nucleon
interaction taken from \cite{Greiner} (for further information see also
\cite{RingSchuck}).

Consider a system consisting of $N_p$ protons and $N_n$ neutrons, which
are described by plain waves:
\begin{equation}
 \varphi_{{\alpha}}=\frac{1}{\sqrt[]{V}}e^{ikx}\chi_S\chi_I.
\end{equation}
Here, $\chi_I$ is the degree of freedom for the isospin, $\chi_S$ for
the spin. The self--consistent Hartree--Fock equations read:
\begin{eqnarray}
 \epsilon_{{\alpha}}\varphi_{{\alpha}}(x) = 
\hat{t}\varphi_{{\alpha}}(x)+\sum_{{\alpha'}}\int d^3y\varphi_{{\alpha}}^{\dagger}(y)   
V(x,y)\left[\varphi_{{\alpha'}}(y)\varphi_{{\alpha}}(x)-\varphi_{{\alpha}}(y)\varphi_{{\alpha'}}(x)\right].
\end{eqnarray}
where $\sum_{{\alpha'}}$ is the sum over all occupied states. The
potential is spin-- and isospin--independent, so the sum can be divided
into a part for protons and one for neutrons:
\begin{equation}
 \sum_{{\alpha'}}...=\sum_{k,s}^{k_F^{(p)}}...+\sum_{k,s}^{k_F^{(n)}}
\end{equation}
As the sum splits up, one gets a separate Hartree--Fock--equation for
neutrons and protons:  
\begin{eqnarray}
 \epsilon_k^{(p)}\varphi_k^{(p)}(x)=\hat{t}\varphi_k^{(p)}(x)
 &+& (2s+1)\sum_{k'}^{k_F^{(p)}}\int
 d^3y[\varphi_{k'}^{(p)*}(y)\varphi_{k'}^{(p)}(y)]V(x,y)\varphi_{k}^{(p)}(x) 
\nonumber \\ 
 &-& 1\cdot\sum_{k'}^{k_F^{(p)}}\int
 d^3y[\varphi_{k'}^{(p)*}(y)\varphi_{k}^{(p)}(y)]V(x,y)\varphi_{k'}^{(p)}(x) 
\nonumber \\  
 &+& (2s+1)\sum_{k'}^{k_F^{(n)}}\int
 d^3y[\varphi_{k'}^{(n)*}(y)\varphi_{k'}^{(n)}(y)]V(x,y)\varphi_{k}^{(p)}(x).
\end{eqnarray}
by taking into account the spin--degrees of freedom $g_s=2s+1$.  The
neutron wave function results from replacing the superscript $(p)$ by
$(n)$.

Next, the interaction is introduced as a Skyrme--type parameterisation:
\begin{equation}
 V(x,y)=\delta^3(x-y)(\frac{1}{6}t_3 n-t_0),
\end{equation}
where $t_0$ describes the attractive two particle nuclear interaction,
whereas $t_3n$ describes the repulsive (and density depended) many--body
interaction which is the dominant one at high nuclear densities. Here,
$n$ denotes the total number density of the system:
\begin{equation}
 n=(N_n+N_p)/V=n_n+n_p.
\end{equation}
The single--particle--energies for protons and neutrons can then be derived to be:
\begin{eqnarray}
 \epsilon_k^{(p)} &=&
 \frac{k_p^2}{2m_p}+(\frac{1}{6}t_3n-t_0)\frac{1}{V}(N_p/2+N_n)
 \nonumber \\ 
 \epsilon_k^{(n)} &=&
 \frac{k_n^2}{2m_n}+(\frac{1}{6}t_3n-t_0)\frac{1}{V}(N_n/2+N_p). 
\end{eqnarray}
The total energy for a homogeneous system is given by:
\begin{equation}
 E_{HF}=\frac{1}{2}\sum_i^{N_p}(\epsilon_i^{(p)}+t_i)+\frac{1}{2}\sum_i^{N_n}(\epsilon_i^{(n)}+t_i)  
\end{equation}
\begin{equation}
 \Longrightarrow
 E_{HF}=N_p\frac{3}{5}\frac{k_p^2}{2m_p}+N_n\frac{3}{5}\frac{k_n^2}{2m_n} 
 +\frac{1}{2}(\frac{1}{6}t_3n-t_0)\frac{1}{V}(N_p^2/2+2N_pN_n+N_n^2/2).
\end{equation}
The next task is to determine the phenomenological parameters $t_0$ and
$t_3$. For $n=n_0=0.16{\ \rm fm^3}$ the total energy shall have a
minimum with $E_{HF}/A=BE=-16$ MeV. From these conditions one arrives at:
\begin{eqnarray}
 t_3 &=& \frac{16E_B}{n_0^2}+\frac{8}{5}\frac{(\hbar
 c)^2}{m_nc^2}\left(\frac{3\pi^2}{2}\right)^{2/3}(n_0)^{-4/3}
 \nonumber \\ 
 t_0 &=& \frac{16E_B}{3n_0}+\frac{16}{15}\frac{(\hbar
 c)^2}{m_nc^2}\left(\frac{3\pi^2}{2}\right)^{2/3}(n_0)^{-1/3}
\end{eqnarray}
\begin{equation}
 \Longrightarrow t_0=1024.1 {\ \rm MeV\ fm^3}, \quad \
 t_3=14600.8 {\ \rm MeV\ fm^6}\quad . \label{t0t3} 
\end{equation} 
With 
\begin{equation}
K(n_0) = \left.9 \frac{d p(n)}{d n}\right|_{n=n_0}\quad ,
\end{equation}
one finds a value of $376.4$ MeV for the nuclear compressibility, which
is less than for the empirical interaction. 

\begin{figure}[!ht]
{\centering \includegraphics[width=0.6\textwidth]
{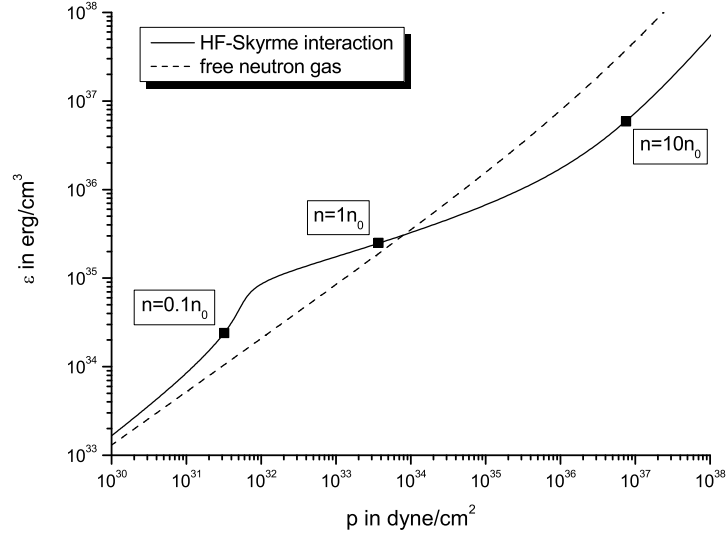}}
\caption{The equation of state for the Skyrme--parameterisation and for a
  free neutron gas.} 
\label{eoshf}
\end{figure}

Again, we assume pure neutron matter, i.e.\ we set $N_p=0$ and $N_n=A$:
\begin{equation}
 \frac{E_{HF}}{A}=\frac{3}{5}\frac{k_n^2}{2m_n}
 +\frac{1}{4}(\frac{1}{6}t_3n-t_0)n. \nonumber \\ 
\end{equation}
With the total energy density $\epsilon(n)$, including the rest mass term $m_n\cdot n$,
\begin{eqnarray}
 \epsilon(n) &=& (m_n+E_{HF}/A)\cdot n \\
 &=& m_nn+\frac{3}{10
 m_n}(3\pi^2\hbar^3)^{2/3}n^{5/3}+\frac{t_3}{24}n^3-\frac{t_0}{4}n^2, 
\label{ehf} 
\end{eqnarray}
one can derive $p(n)$ according to the thermodynamic relation:
\begin{eqnarray}
 p(n) &=& n \frac{d}{dn}\epsilon(n) - \epsilon(n) \\
 \Longrightarrow p(n) &=& \frac{2}{10
 m_n}(3\pi^2\hbar^3)^{2/3}n^{5/3}+\frac{t_3}{12}n^3-\frac{t_0}{4}n^2. 
\end{eqnarray}
The resulting equation of state is plotted in Fig.~\ref{eoshf}, which
appears to be very similar to the empirical interaction discussed before.

\begin{figure}[!ht]
{\centering \includegraphics[width=0.6\textwidth]
{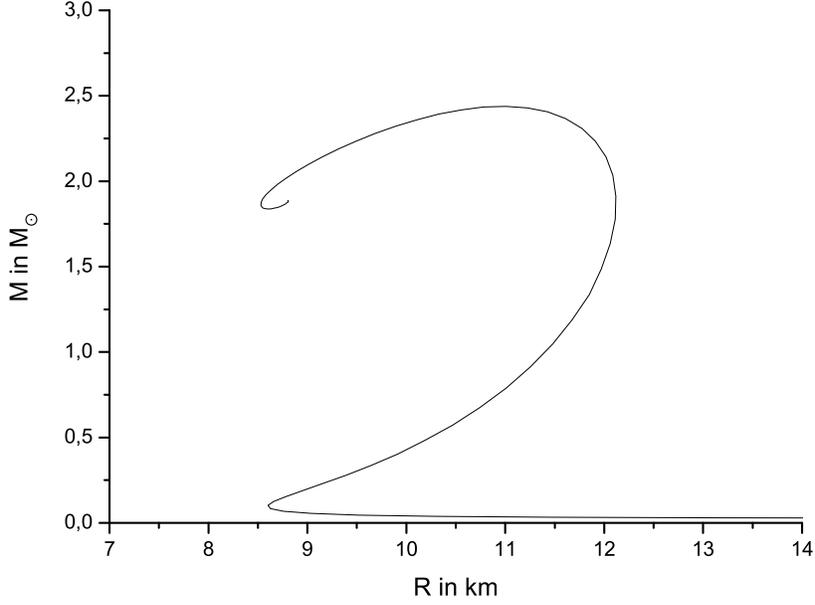}}
\caption{The mass $M$ as a function of $R$ for the
  Skyrme--parameterisation. The maximum mass star has a mass of $M=2.438
  {\ \rm M_{\odot}}$ and a radius of $R=10.98$ km.} 
\label{mrhf}
\end{figure}

Fig.~\ref{mrhf} depicts the corresponding mass--radius plot. The maximum
mass star has a mass of $M=2.438 {\ \rm M_{\odot}}$ and a radius of
$R=10.98$ km. These values are only a bit smaller than the ones
calculated before for the empirical interaction, also the shape of the
curve looks similar. The sharp turnover in the lower left corner of the plot
corresponds to a minimal radius of only $R\simeq8.6$ km and a mass of
$M\simeq0.1{\ \rm M_\odot}$.

\subsection{Comparing the two models}

It is instructive to take a closer look on the two different EoSs
considered here, as the results do not differ that much from each other.
The two energy densities (eqs.~(\ref{eemp}) and (\ref{ehf})) are:
\begin{itemize}
\item empirical nucleon--nucleon parameterisation: 
\begin{equation} 
\frac{\epsilon_{emp}(n)}{n} = m_n + \left<E_0\right> u^{2/3} + \frac{A}{2} u 
 + \frac{B}{\sigma + 1} u^\sigma +(2^{2/3} - 1)
 \left<E_0\right>\left( u^{2/3} - u \right) + S_0 u
\end{equation}
\item Hartree--Fock method and Skyrme parameterisation:
\begin{equation} 
\frac{\epsilon_{HF}(n)}{n}=m_n+\frac{3}{10 m_n}(3\pi^2\hbar^3)^{2/3}n^{2/3}+
 \frac{1}{4}(\frac{1}{6}t_3n-t_0)n
\end{equation} 
\end{itemize}
With $n=u\cdot n_0$ and eq.~(\ref{e0}) one can rewrite the two
equations in the following form: 
\begin{eqnarray}
 \frac{\epsilon_{emp}(n)}{n} &=& m_n+ 2^{2/3}
 \left<E_0\right>u^{2/3} +\left(\frac{A}{2} -(2^{2/3} - 1)
 \left<E_0\right>+S_0\right)u + 
 \frac{B}{\sigma + 1} u^\sigma \\
 \frac{\epsilon_{HF}(n)}{n} &=& m_n + 2^{2/3}\left<E_0\right>
 u^{2/3} - \frac{1}{4}t_0n_0u 
 +\frac{1}{24}t_3n_0^2u^2.
\end{eqnarray}
By inserting the numbers (\ref{t0t3}), the energy density exhibits the
following parametric form: 
\begin{eqnarray}
 \frac{\epsilon_{emp}(n)}{n} &=& (939.6+ 35.1u^{2/3}-42.1u +21.0
 u^{2.112}) {\ \rm MeV} \stackrel{u\gg1}{\longrightarrow} 21.0
 u^{2.112} {\ \rm MeV}\\ 
 \frac{\epsilon_{HF}(n)}{n} &=& (939.6+
 35.1u^{2/3}-41.0u+15.6u^2) {\ \rm MeV}
 \stackrel{u\gg1}{\longrightarrow} 15.6u^2 {\ \rm MeV}. 
\end{eqnarray}
If the energy density follows a power--law as a function of $n$, it is
easy to find an explicit equation of state of the form $p(\epsilon)$: 
\begin{eqnarray}
 \epsilon &=& c\cdot n^k \nonumber \\
 p &=& n\cdot \frac{d\epsilon}{dn}-\epsilon=(k-1)c\cdot n^k \nonumber \\
 p(\epsilon) &=& (k-1) \epsilon \label{pke}\quad .
\end{eqnarray}
In our case, $k=3.112$ for the empirical interaction and $k=3$ for the
Skyrme--parameterisation. One can see, that the first EoS is stiffer than
the second. This explains why the masses in the empirical model are
bigger as the masses in the Skyrme--parameterisation, the harder EoS can
support more mass. 

\begin{figure}[!ht]
{\centering \includegraphics[width=0.6\textwidth]
{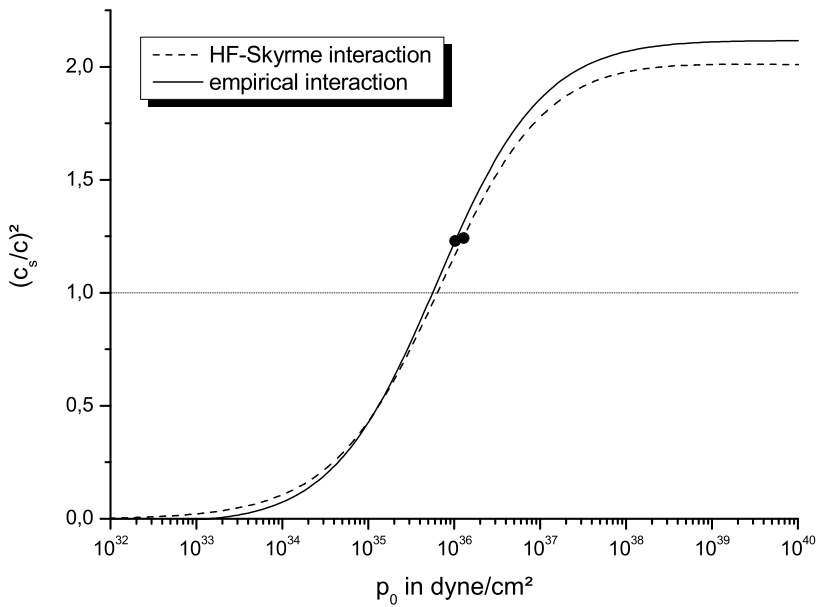}}
\caption{The speed of sound depending on the central pressure for the
  two different models. The maximum mass stars are marked with a dot.}
\label{sos}
\end{figure}

\subsection{Speed of sound and causality}

The speed of sound in our neutron star must be less than the speed of
light to fulfil relativistic causality. The adiabatic speed of sound is
given by $c_s=\frac{dp}{d\epsilon}$ in a non--dispersive medium. Hence,
we have:
\begin{equation}
 1\leq\left(\frac{c_s}{c}\right)^2=\frac{dp}{d\epsilon}
=\frac{dp/dn}{d\epsilon/dn}\quad .  
\end{equation}
Using eq.~(\ref{pke}) one gets:
\begin{equation}
 \frac{dp}{d\epsilon} = k-1 \leq 1.
\end{equation}
Hence, the power coefficient $k$ must be less than or equal two. So both
EoSs will become acausal for large $n$. This behaviour is demonstrated in
Fig.~\ref{sos}, where the speed of sound is plotted against the central
pressure. As one can see, for the maximum mass star the speed of sound
$c_s/c$ is larger than one for both interactions considered.  hence,
correction factors have to be added to the interaction, which include
(or simulate) relativistic effects (see e.g.\ \cite{Akmal98}).

\section{Summary and brief overview on present day developments}

In chapter \ref{whitedwarfs}, we discussed the common structure equations 
for stars as well as the
relativistic and non--relativistic equations of state (EoS) for a Fermi
gas of electrons and nucleons. We simplified the EoS to a polytropic
form and derived semi--analytically the Chandrasekhar mass limit of M =
$1.4312 \left(2/\eta \right)^2 {\rm M_\odot}$ for white dwarfs. We
calculated numerically mass--radius relations for white dwarfs using the
relativistic and non--relativistic polytropic approximations. For the
relativistic case we find the computed masses to be constant and in very
good agreement with the Chandrasekhar mass limit. The numerical results
were also checked by the analytical results for white dwarf radii.

We introduced a numerical method to calculate the EoS of a free gas of
particles for the general case, i.e.\ without using the polytropic
approximation, and computed the corresponding masses and radii for white
dwarfs. In this case, the calculated masses became larger with
increasing central pressures up to a maximum mass of around 1.4 ${\rm
  M_\odot}$. For high pressures, the masses stayed constant at the
maximum value showing that the EoS turned into a relativistic polytrope.

We calculated the mass--radius relations without as well as with
corrections from general relativity. The mass curve showed an increasing
mass for larger central pressures up to $p_0 \sim 10^{28} {\rm
  dyne/cm^2}$ reaching a maximum mass of about 1.40 ${\rm M_\odot}$
without and one of 1.39 ${\rm M_\odot}$ with relativistic corrections.
The slightly reduced maximum mass in the latter case was in accord with
the fact that general relativity corrections strengthen the
gravitational force.

We discussed also that white dwarfs and neutron stars are stable against
radial oscillations for masses smaller than the maximum mass but
unstable beyond the mass peak.

In chapter \ref{ns}, we looked at pure neutron stars using an EoS of a
free Fermi gas of neutrons. We computed the exact equation of state by a
root finding routine. The maximum mass for such a neutron star was found
to be $M=0.712$ M$_\odot$ with a radius of $R=9.14$ km. Then, we
included protons and electrons in $\beta$--equilibrium to the
composition of neutron star matter. The maximum mass of the
corresponding neutron star changed slightly to $M=0.699{\ \rm M_\odot}$
and the radius to $R=9.24$ km.

Next, we focused on the impact of nuclear interactions on the properties
of pure neutron stars. We analysed two different types of interaction,
an empirical interaction and the Hartree--Fock method applied to a
Skyrme--parameterisation. As the two resulting EoSs were very close to
each other, the results for the mass and radius curves were very
similar, but quite different to the cases of free Fermi gases considered
before. The maximum masses of neutron stars reach now values of
$M=2.4M_\odot$ and $M=2.8M_\odot$, respectively, which are up to four
times larger than for the free cases. The corresponding radii increase
to $R=11$ to 12 km, respectively. The repulsion between nucleons
stiffens the EoS, increases the pressure at fixed energy density, and
enables to stabilise more massive neutron stars compared to the free
Fermi gas case. Caveats of the simple nuclear interaction models used
were exhibited, as the neglect of relativistic effects and the EoSs
becoming acausal for large central pressures.

We close this section with a brief discussion of present day
developments in the field of compact stars: astrophysical data, the EoS
of dense matter and its impact on the global properties of compact stars
(for a recent general introduction on latest developments in the field
we refer to \cite{LP04}).

More than 1500 pulsars, rotating neutron stars with pulsed radio
emission, are known today. About seven thermally emitting isolated
neutron stars have been observed by the Hubble Space Telescope (HST), by
the x--ray satellites Chandra and XMM--Newton and by ground based
optical telescopes as the European Southern Observatory ESO. For the
closed and best studied one, RX J1856.5-3754, the measured spectra can
be well described by a Planck curve \cite{Drake02}. As the outermost
layer of a neutron star is the atmosphere up to a density of $10^4$
gm/cm$^3$ consists of atoms, their presence should show up in the
spectra. However, there is a lack of any atomic spectral line in the spectra
which is not understood at present \cite{Burwitz03}.  For three selected
rotating neutron stars (pulsars), nicknamed the Three Musketeers, one is
even able to make phase resolved studies of the spectra, which allows to
determine that there are hot spots on the surface of the neutron star
\cite{DeLuca05}.

The high--density EoS for neutron stars and the appearance of exotic
matter has been studied in modern relativistic field--theoretical models
\cite{Glen_book}. In particular, hyperons will appear around twice
normal nuclear matter density (see e.g.~\cite{JSB04} and references
therein). The possibility of Bose condensation of kaons in dense, cold
matter has been investigated also \cite{Li97}.  Last but not least, the
phase transition to strange quark matter, quark matter with strange
quarks, will appear in the high--density limit \cite{Weber05,JSBsqm04}.
If the phase transition of these exotic forms of matter is strongly
first order, a new class of compact stars emerges as a new stable
solution of the TOV equations.  While the existence of this third family
of compact stars besides white dwarfs and neutron stars was known
earlier \cite{Gerlach68,Haensel80,Kaempfer81}, the solution was
rediscovered in modern models for the EoS and a Gibbs phase transition
only recently \cite{GK2000,Schertler00,Scha02,FPS01}. The topic of the
third family of compact stars was a subject for another undergraduate
student project of Jean Macher and one of the authors \cite{Macher05}.
The study of dense quark matter, as possibly present in the core of
neutron stars, has been the object of intense investigation during the
last few years due to the rediscovery of the phenomenon of colour
superconductivity. Here, two colour charged quarks form a (colour)
superconducting state and change the properties of quark matter (for
reviews see e.g.~\cite{KrishnaReview,MarkReview,DirkReview}). Neutron
stars with colour superconducting quark matter have been calculated by
many authors in the last couple of years, see
\cite{Alford03,Lugones03,Baldo03,Banik03,Blaschke03,Shovkovy03,Grigorian04,Ruster04,Buballa04,Drago04,Alford04}.
The physics of strange quark matter in compact stars and its possible
signals are reviewed most recently also in \cite{Weber05,JSBsqm04}.

The physics of neutron star has also impacts not only on pulsars but
also in other active research fields of astrophysics. For example, in
x--ray binaries a neutron star accretes matter from his companion, which
is an ordinary star or a white dwarf. When the matter falls onto the
neutron star, x--ray bursts occur. Redshifted spectral lines have been
measured from such x--ray binaries, which allow to constrain the
compactness and by that the EoS of neutron stars \cite{Cottam02}.  Core
collapse supernovae form a hot proto--neutron star first which cools
subsequently and emits a neutrino wind. That neutrino wind influences the
production of heavy elements in the supernova material in the so called
r--process nucleosynthesis (see e.g. \cite{Langanke03} and references
therein). Neutron stars can collide with a corresponding emission of
gravitational waves, which will be measured by gravitational wave
detectors in the near future, and gamma--rays, being considered a prime
candidate for gamma-ray bursts and another site for r--process
nucleosynthesis. The dynamics of the collision depends on the
underlying EoS of neutron stars \cite{Oechslin04}! In the future, the
square kilometre array (SKA) will measure more than 10.000 pulsars, with
expected 100 binary systems with a neutron stars \cite{Kramer03}, so the
prospects are bright for learning more about the physics of neutron
stars!

\bibliography{all,literat,proseminar}

\end{document}